\documentclass[fp,twocolumn]{jpsj3}
\usepackage{txfonts}
\usepackage{bm}
\usepackage[dvipdfmx]{graphicx}
\usepackage[dvipdfmx]{color}
\usepackage{amsmath, amssymb}
\usepackage{braket}
\usepackage{color}
\usepackage{comment}

\title {Theoretical perspectives on optical control of magnetism in spin-charge coupled systems}
\author{Masahito Mochizuki\thanks{masa{\_}mochizuki@waseda.jp}}
\inst{Department of Applied Physics, Waseda University, Okubo, Shinjuku-ku, Tokyo 169-8555, Japan}
\abst{In this article, we review recent theoretical research on the photocontrol of magnetism in several types of spin-charge coupled systems. The control of magnetism with light has been a central issue in condensed-matter physics, attracting enormous research interest both for fundamental science and for technological applications. This field of research has developed rapidly in recent years along with the development of laser technology. However, because the direct coupling between the  light magnetic field and magnetization via the Zeeman coupling is very weak in terms of the energy scale, it is, in principle, difficult to induce dramatic effects as far as this magnetic light-matter interaction is exploited. On the contrary, the interaction between the light electric field and electron charges has an energy scale two to three orders of magnitude larger than that of the magnetic interaction. Therefore, we may realize astonishing photoinduced physical phenomena and novel optical device functions by exploiting this electric light-matter interaction. Spin-charge coupled magnets, e.g., double-exchange magnets, multiferroics materials, and Rashba electron systems, in which spins and charges are strongly coupled through several kinds of mechanisms such as exchange interactions and spin-orbit coupling, are ideal systems for realizing this idea. Recent theoretical studies have revealed that it is possible to control, manipulate and switch the magnetization coupled to electron charges in these systems through exciting and/or driving them with  light electric fields. The following three recent topics are discussed as examples of such theoretical studies, that is, photoinduced magnetic phase transitions in irradiated double-exchange models, highly efficient photoinduction of spin polarization in Rashba electron systems, and electromagnon excitations and their intense excitation effects in multiferroic materials.}

\begin{document}
\maketitle
\section{Introduction}
\begin{figure*}[tb]
\centering
\includegraphics[scale=0.5]{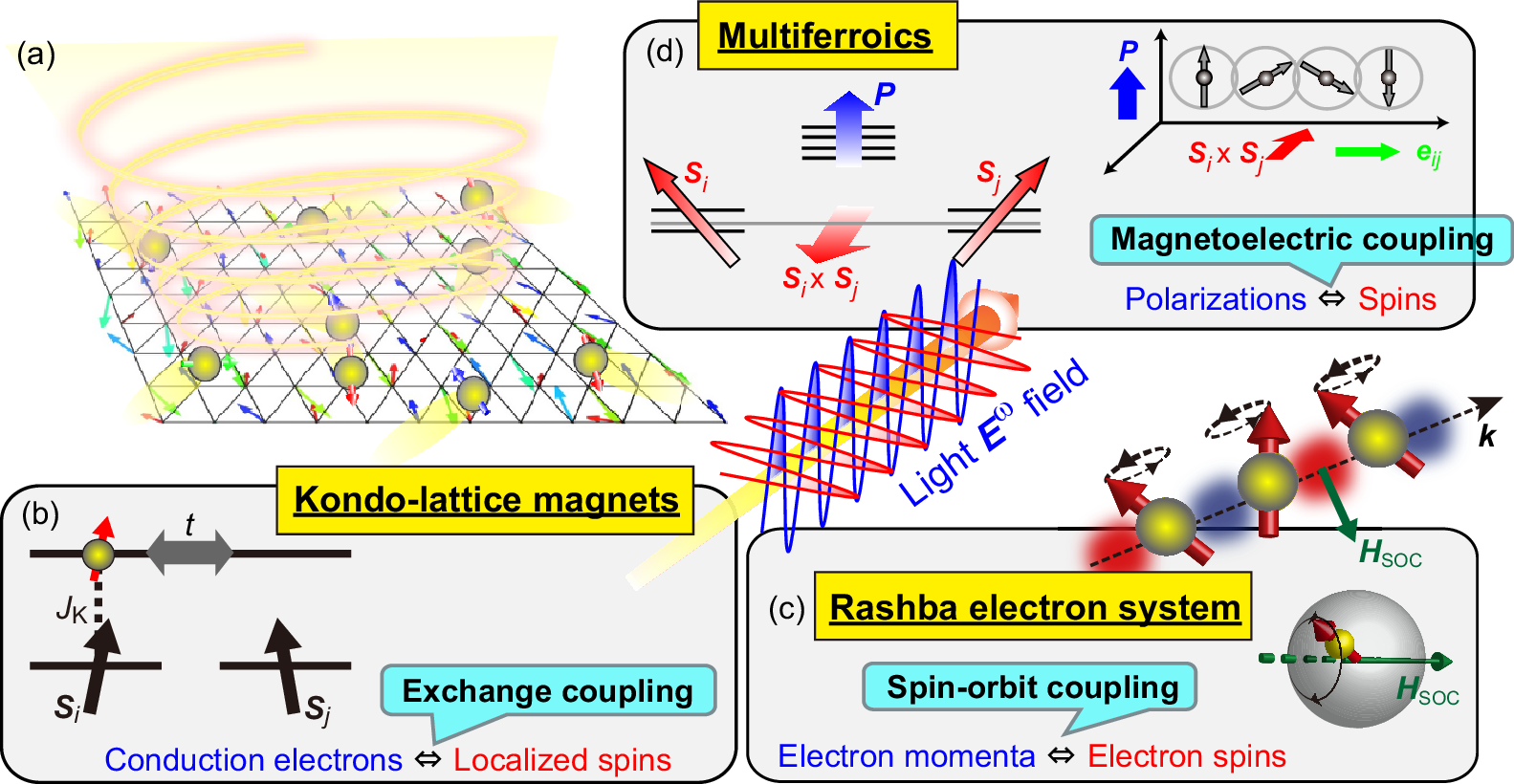}
\caption{(a) Schematics of the photocontrol of magnetism in spin-charge coupled systems. (b)-(d) Examples of the spin-charge coupled systems, i.e., (b) double-exchange magnet with exchange coupling between conduction electrons and localized spins, (c) Rashba electron system with spin-orbit coupling between electron momenta and electron spins, and (d) Spiral-magnetism-based multiferroic material with magnetoelectric coupling via the antisymmetric exchange striction mechanism.}
\label{Fig01}
\end{figure*}
Control of magnetism with light has attracted enormous research interest from the perspective of fundamental science for its diverse physical mechanisms with light-matter interactions [see Fig.~\ref{Fig01}(a)]~\cite{Kirilyuk2010}. It also has many advantages in terms of technical applications, offering us high-speed, energy-saving, and non-contact techniques to control magnetization. Thus, the optical control of magnetism has been intensively studied, and this research field has been developed rapidly along with the recent development of laser technology. In this field, research has so far been carried out on the demagnetization~\cite{Beaurepaire1996} and generation~\cite{Koshihara1997} of ferromagnetic magnetization by light irradiation, and on the generation of simple magnetic structures such as magnetic vortices by light~\cite{Ogasawara2009,Finazzi2013}. However, there have been few examples of studies on the generation and switching of complex noncollinear magnetic long-range orders by light. On the other hand, recent studies have revealed that noncollinear magnetic orders, such as spiral magnetism and skyrmion crystals, are treasure troves of diverse physical phenomena and material functions originating from magnetoelectric coupling and magnetic topology~\cite{Nagaosa2012,Nagaosa2013,Tokura2014,Tokura2017}. The generation, switching and excitation of long-range orders of such topological and noncollinear magnetism are important for both fundamental science and technical applications, but have not necessarily been successful.

In fact, the following difficulties exist in the control of magnetism by light, (1) Magnetization dynamics are slow in time scale and thus can only interact with low-energy electromagnetic waves (light and microwaves) directly. (2) The interaction between magnetization and light magnetic field is so small in energy scale that dramatic effects and phenomena such as phase transitions and switching are difficult to achieve. (3) The magnetic orders in localized magnets are largely determined by spin-spin exchange interactions inherent in the materials, which are hard to be modulated by light. Hence, it is difficult to induce dramatic effects and phenomena as far as direct interactions between the light magnetic field and magnetization are exploited.

These difficulties may be overcome by utilizing spin-charge coupled magnets, in which the localized spins are strongly coupled to conduction electrons through exchange interactions or spin-orbit interactions as the target materials. More specifically, spin-charge coupled magnets with photoactive electron systems have the following three advantages, which can solve each of the above three difficulties. (1) Charge dynamics are six to nine orders of magnitude faster than magnetization dynamics so that they can interact directly with electromagnetic waves of higher energy. (2) Interactions between light electric fields and conduction-electron charges have two to three orders of magnitude higher energy than interactions between light magnetic fields and localized spins so that dramatic phenomena and effects can be expected. (3) Through changing the electronic structures such as band dispersions, band width, Fermi-surface structure, and density of states with light, the exchange interactions among localized spins, which are mediated by conduction electrons, can be modulated significantly.

Let us discuss the second advantage in detail. According to the electromagnetism, the relation $E_0/B_0=c$ holds for electromagnetic waves where $E_0$ and $B_0$ are amplitudes of the electric-field and magnetic-field components, respectively, and $c$ is the speed of light. The energy associated with the magnetic-field component comes from the Zeeman coupling to the electron spins, whose magnitude is typically $E_{\rm mag}=g\mu_{\rm B}SB_0$ where $S(=1/2)$ is the spin quantum number of electron. On the other hand, the energy associated with the electric-field component comes from coupling to the electron dipoles, which is typically $E_{\rm ele}=exE_0$ where $e$ is the elementary charge. Here $x$ is a typical length scale of the displacement of charge distribution. When we consider electric dipoles originating from spatially imbalanced charge distribution on the atoms, we should adopt the Bohr radius $a_{\rm B}$ for $x$. In this case, the ratio of $E_{\rm ele}/E_{\rm mag}$ becomes an inverse of the fine-structure constant, i.e., $\sim 137$. On the contrary, when we consider electric dipoles originating from the crystallographic displacement of atoms, $x$ is typically a lattice constant $a$ of the crystal. Considering that $a$ is typically four to seven times larger than $a_{\rm B}$, the order of the ratio $E_{\rm ele}/E_{\rm mag}$ is $10^3$. Thus, we can conclude that the electric light-matter interaction with light electric field has two to three orders of magnitude stronger than the magnetic light-matter interaction with light magnetic field.

Based on this idea, theoretical studies have been done exploring novel photoinduced phenomena in spin-charge coupled systems such as double-exchange magnets, Rashba electron systems, and multiferroic materials [Figs.~\ref{Fig01}(b)-(d)]. Corresponding materials have been successively discovered and synthesized in recent years, which have further accelerated the research. In these studies, numerical simulations of mutually coupled spatiotemporal dynamics of mutually coupled spin and electron systems have been successful, and various interesting phenomena have been discovered and predicted. In addition, there has been active research through theoretical analysis based on Floquet theory, which maps a problem of the nonequilibrium steady states driven by a time-periodic external field onto an effective static problem of equilibrium states. These studies have elucidated many fundamental and universal principles behind the photoinduced magnetic phenomena and proposed experimental guidelines for realizing and observing novel phenomena and functions.

In this article, we will introduce several examples of such recent theoretical studies. Specifically, the following three topics will be discussed.\\
\\
\noindent
{\bf [1] Theoretical study of photoinduced magnetic phase transitions in double-exchange models~\cite{Inoue2022}:}
The double-exchange models (or the Kondo-lattice models) describe a system in which localized spins are coupled to conduction electrons through exchange interactions~\cite{Zener1951,Anderson1955,Gennes1960}. It is known that Ruderman-Kittel-Kasuya-Yoshida (RKKY) type long-range interactions among the localized spins mediated by conduction electrons give rise to long-period noncollinear magnetic orders with magnetic modulation vectors governed by the nesting of Fermi surfaces. We discuss the emergence of photoinduced 120-degree spin order in the double-exchange model on a triangular lattice~\cite{Inoue2022}. Numerical simulations reveal that this phenomenon is caused by a peculiar electron-occupation state under photoirradiation called pseudo half-filling state, which is realized via photoexcitation of conduction electrons, formation of a band gap due to the dynamical localization effect, and relaxation of the excited electrons to the lower band. This physical mechanism is universal and expected not only in triangular double-exchange models but also in the models on several types of lattices.\\
\\
\noindent
{\bf [2] Microscopic theory of highly efficient photoinduction of spin polarization in Rashba electron systems~\cite{Mochizuki2018,Tanaka2020}:}
In systems with broken spatial inversion symmetry, the spin-orbit interaction arises and links the momentum and spin of conduction electrons~\cite{Rashba1960,Dresselhaus1955,WinklerBook2003,Manchon2015}. This momentum-spin coupling induces an effective magnetic field acting on the electron when it is driven by an external electric field. Importantly, the direction of the effective magnetic field depends on the direction of motion of the electron. Therefore, circular motion of conduction electrons induced by circularly polarized light electric field interacting with the electron charge gives rise to an effective rotational magnetic field and thus circulation of the electron spin. The angular momentum of the spin circulation is converted to the spin polarization perpendicular to the circularly polarized plane. Numerical simulations show that in magnetic semiconductors and heterojunction systems with strong Rashba spin-orbit interactions, irradiation with circularly polarized light can induce ferromagnetic spin polarization with extremely high efficiency~\cite{Mochizuki2018,Tanaka2020}. Furthermore, the analysis using the Floquet theory has comprehensively reproduced the sign change of the photoinduced spin polarization depending on the electron filing found in the simulations as well as the dependence of the induced spin polarization on the light intensity and frequency.\\
\\
\noindent
{\bf [3] Theory of electromagnon excitations in multiferroics and predicted their intense excitation effects~\cite{Mochizuki2010a,Mochizuki2010b,Mochizuki2011a}:}
Spiral magnetism in insulating magnets causes an imbalance of electron-charge distribution through spin-orbit interactions and thus induces ferroelectric polarization, which realizes the magnetism-based ferroelectric system called multiferroics in which ferroelectric and magnetic orders coexist~\cite{Tokura2014}. In such a system, oscillations of the electric polarizations induced by light electric fields can cause oscillations of the spins via their mutual coupling called magnetoelectric coupling. This combined excitation of electric polarizations and spins is called electromagnon. The interaction between the light electric field and the electric polarizations is two or three orders of magnitude larger than the interaction between the light magnetic field and the spins. Therefore, intense excitation is possible for electromagnons in contrast to ordinary magnons. Intensely excited electromagnons are expected to exhibit high nonlinearity and dramatic phenomena. As an example of such phenomena, we will discuss the physical mechanism of electromagnon excitations in multiferroic perovskite manganites $R$MnO$_3$ ($R$=Tb, Dy, Eu$_{1-x}$Y$_x$, etc.)~\cite{Mochizuki2010a} and the theoretical prediction of optical switching of the vector spin chirality through intense electromagnon excitations in this class of materials~\cite{Mochizuki2010b,Mochizuki2011a}.

\section{Photoinduced magnetic phase transitions in double-exchange models}
We first discuss a theoretical study on the photoinduced magnetic phase transitions in double-exchange models by taking a triangular double-exchange model as an example~\cite{Inoue2022}. The double-exchange model is a lattice spin-electron model describing a complex system of conduction electrons and localized spins which are mutually coupled via exchange interactions~\cite{Zener1951,Anderson1955,Gennes1960}. This model and related models provide a precious platform for studying dramatic photoinduced spin-charge coupled phenomena~\cite{Inoue2022,Koshibae2009,Koshibae2011,Ono2017,Ono2018,Ono2019,Eto2024b,Hattori2024,Hamano2025,Matsueda2007,Kanamori2009,Ohara2013}. The Hamiltonian of the double-exchange mode at equilibrium is composed of two terms, the first term that describes kinetic energies of the conduction electrons and the second term that describes the exchange interactions between the localized spins and the conduction-electron spins, which is given by,
\begin{align}
\label{eq:Eq01}
\mathcal{H}=\sum_{\langle i,j \rangle,\sigma} t_{ij}c^\dag_{i\sigma}c_{j\sigma}
-J_\mathrm{K}\sum_{i,\sigma,\sigma'}\bm{S}_i \cdot c^\dag_{i\sigma}\bm{\sigma}
_{\sigma\sigma'}c_{i\sigma'}.
\end{align}
In the first term, only the nearest-neighbor hoppings are considered on a triangular lattice, i.e., $t_{ij}=-t$ for the adjacent site pairs $i$ and $j$, while $t_{ij}=0$ for other site pairs. Here $(c_{i\sigma})$and $c^\dag_{i\sigma}$ are annihilation and creation operators of a conduction electron with spin $\bm \sigma$ on the $i$th site, respectively. In the second term, the symbol $\bm \sigma \equiv (\sigma_x, \sigma_y, \sigma_z)$ is a vector of the Pauli matrices describing the conduction-electron spins, while $\bm S_i$ is a normalized classical vector ($|\bm S_i|=1$) describing the localized spins. Here we consider the ferromagnetic exchange coupling $J_{\rm K}\,(>0)$. The electron filling $n_{\rm e}$ is defined by,
\begin{align}
\label{eq:Eq02}
n_{\rm e} \equiv \frac{1}{2N}\sum_{i,\sigma}\braket{c_{i\sigma}^\dagger c_{i\sigma}},
\end{align}
where $N$ is the number of lattice sites.

\begin{figure}[tb]
\centering
\includegraphics[scale=0.5]{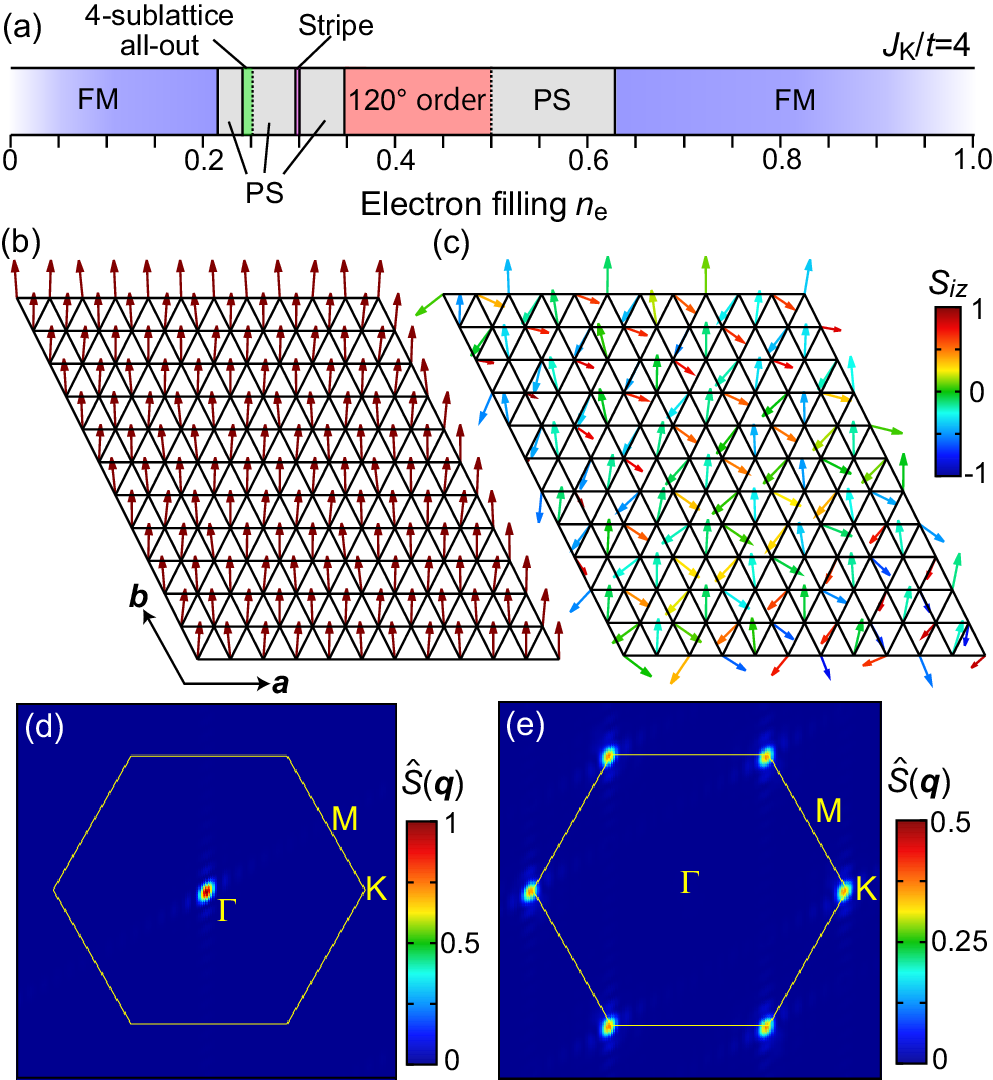}
\caption{(a) Ground-state phase diagram of the triangular double-exchange model with $J_{\rm K}/t=4$ as a function of the electron filling $n_{\rm e}$. (b),(c) Spin configurations of the ferromagnetic state at equilibrium (b) and the 120-degree spin state in the photodriven system (c). (d),(e) Spin structure factors $\hat{S}(\bm q)$ for these magnetic states, where the hexagons represent the first Brillouin zone. This figure is taken and modified from Ref.~\cite{Inoue2022} {\copyright} 2022 American Physical Society.}
\label{Fig02}
\end{figure}
The effects of light irradiation are considered through the Peierls phases attached to the transfer integrals $t_{ij}$ as,
\begin{align}
\label{eq:Eq03}
t_{ij} \; \rightarrow \;
t_{ij}\exp\left[-i\bm A(\tau) \cdot (\bm r_i - \bm r_j)\right],
\end{align}
where $\tau$ and $\bm r_i$ are time and spatial coordinates of the $i$th site, respectively. With the radiation gauge, the time-dependent vector potential $\bm A(\tau)$ is related with a light electromagnetic field $\bm E(\tau)$ as,
\begin{align}
\label{eq:Eq04}
\bm E(\tau)=-\frac{\partial\bm A(\tau)}{\partial \tau}.
\end{align}
Hereafter natural units $e=\hbar=c=1$ are used, and the transfer integral $t$ and the lattice constant $a$ are adopted as the units of energy and length, respectively. The unit conversions for several physical quantities when $t=1$ eV and $a=5$ \AA are summarized in Table~\ref{tab:unitconv}.
\begin{table}[tb]
\caption{Unit conversions when $t$=1 eV and $a$=5 \AA.}
\centering
\begin{tabular}{lll}
\hline Qunantity & Dimensionless quantity & Corresponding value \\
\hline \hline
Frequency & $\omega = \hbar\tilde{\omega}/t=1$ & $f=\tilde{\omega}/2\pi=242$ THz \\
$E$ field & $E^\omega=ea\tilde{E}^\omega/t=$1 & $\tilde{E}^\omega=20$ MV/cm \\
Time & $\tau=\tilde{\tau}t/\hbar=$1 & $\tilde{\tau}=0.66$ fs \\
\hline
\end{tabular}
\label{tab:unitconv}
\end{table}
The triangular double-exchange model in Eq.~\eqref{eq:Eq01} hosts several magnetic states as shown by its ground-state phase diagram in Fig.~\ref{Fig02}(a)~\cite{Akagi2010,Akagi2012,Azhar2017}. This phase diagram indicates that for a typical strength of the Kondo coupling of $J_{\rm K}=4t$, the long-range 120-degree spin order appears at and below the half filling, i.e., $0.35 \leq n_{\rm e} \le 0.5$, while the ferromagnetic order appears in the lower-filling region of $n_{\rm e} \leq 0.22$ and the higher-filling region of $n_{\rm e} \geq 0.63$ [Figs.~\ref{Fig02}(b)-(e)].

Mutually coupled dynamics of the conduction electrons and the localized spins in the photoirradiated system are simulated by numerically solving the time-dependent Schr\"odinger equation and the Landau-Lifshitz-Gilber equation simultaneously. The former equation describes the spatiotemporal evolution of one-particle wavefunctions $\ket{\tilde{\Psi}_\mu(\tau)}$ for the conduction electrons, which is given by,
\begin{align}
\label{eq:Eq05}
i\partial_\tau\ket{\tilde{\Psi}_\nu(\tau)} = \mathcal{H}(\tau)\ket{\tilde{\Psi}_\nu(\tau)},
\end{align}
where $\nu$ ($=1, 2,\ldots, N_{\rm e}$) is a label of the one-particle states. For the conduction-electron system, the wave function at time $\tau$ is given by an antisymmetric Cartesian product of the one-particle states as,
\begin{align}
\label{eq:Eq06}
\ket{\tilde{\Phi}(\tau)}=\ket{\tilde{\Psi}_1(\tau)} \otimes \ket{\tilde{\Psi}_2(\tau)} \otimes \cdots \otimes \ket{\tilde{\Psi}_{N_\mathrm{e}}(\tau)}.
\end{align}

On the other hand, the latter equation describes the spatiotemporal evolution of localized spins $\bm S_i(\tau)$, which is given by,
\begin{align}
\label{eq:Eq07}
\partial_\tau\bm{S}_i= \bm{h}^{\mathrm{eff}}_i\times\bm{S}_i
+ \alpha_{\rm G}\bm{S}_i\times\partial_\tau\bm{S}_i,
\end{align}
where $\alpha_{\rm G}$ is the Gilbert-damping constant which represents strength of the phenomenologically introduced dissipation effects. Note that the Gilbert damping depicts the energy dissipation of not only the localized-spin system but also the conduction-electron system coupled to the former system via the Kondo coupling.

These two equations are coupled via the effective local magnetic field $\bm{h}^{\rm eff}_i$. This effective field is generated by conduction electrons and act on the localized spin $\bm S_i$ via the Kondo coupling. By the mean-field decoupling of the Kondo-coupling term, its expression is given by,
\begin{align}
\label{eq:Eq08}
\bm{h}^{\rm eff}_i 
=-\left\langle \frac{\partial\mathcal{H}(\tau)}{\partial\bm{S}_i} \right\rangle 
=J_\mathrm{K}\sum_{\sigma,\sigma'}
\braket{\tilde{\Phi}(\tau)|c^\dag_{i\sigma}
\bm \sigma_{\sigma\sigma'} c_{i\sigma'}|\tilde{\Phi}(\tau)}
\end{align}
with
\begin{align}
\label{eq:Eq09}
\braket{\tilde{\Phi}(\tau)|c^\dag_{i\sigma}
\bm \sigma_{\sigma\sigma'} c_{i\sigma'}|\tilde{\Phi}(\tau)}
=\sum_{\alpha=1}^{N_{\rm e}} \bm \sigma_{\sigma\sigma'} 
\braket{\tilde{\Psi}_\alpha(\tau)|i\sigma}
\braket{i\sigma'|\tilde{\Psi}_\alpha(\tau)}.
\end{align}
For more details of the numerical method, see Refs.~\cite{Ono2017,Inoue2022}.

The numerical simulations demonstrate that light irradiation induces a dynamical phase transition from the ground-state ferromagnetic state to the 120-degree spin state as a nonequilibrium steady state in the photodriven system. Figures~\ref{Fig02}(b) and (c) show real-space configurations of the localized spins in the ferromagnetic state before the photoirradiation and the 120-degree spin state under the photoirradiation, respectively. Corresponding spin structure factors $\hat{S}(\bm q)$ in the momentum space are shown in Figs.~\ref{Fig02}(d) and (e), respectively. Note that the spin structure factor $\hat{S}(\bm q)$ has a large peak at $\Gamma$ point in the ferromagnetic state, while it has larger peaks at K point in the 120-degree spin state.

\begin{figure}[tb]
\centering
\includegraphics[scale=0.5]{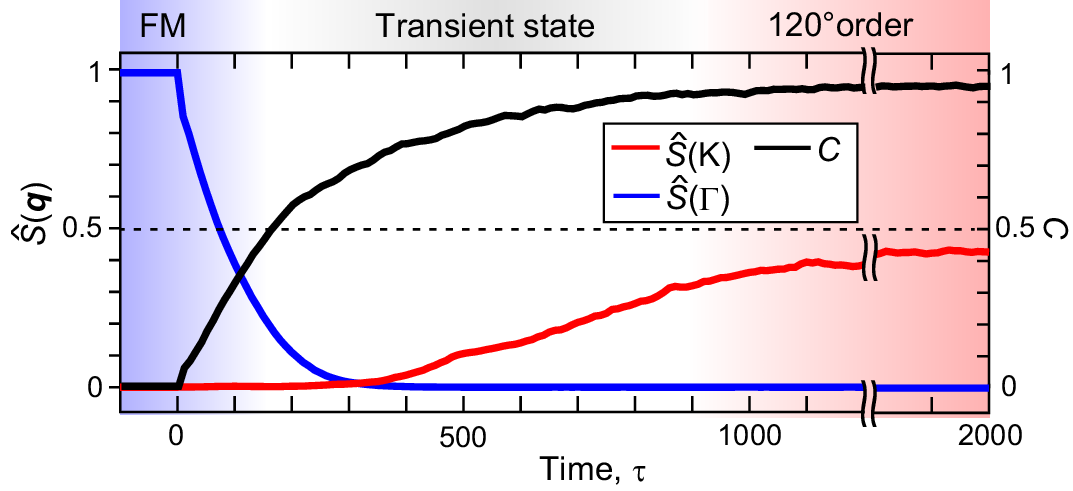}
\caption{Simulated time evolutions of $\hat{S}(\bm q)$ at $\Gamma$ and K points and the averaged vector spin chirality $C=\sum_i |\bm C_i|/N$ in the process of the photoinduced magnetic phase transition from ferromagnetic (FM) to 120-degree spin ordered states. The vector spin chirality $C$ is normalized such that it takes unity for a perfect (single-domain) 120-degree spin order. For parameter values and the system size used for the simulations, see Ref.~\cite{Inoue2022}. This figure is taken and modified from Ref.~\cite{Inoue2022} {\copyright} 2022 American Physical Society.}
\label{Fig03}
\end{figure}
For the photoinduced phase transition from the ferromagnetic state to the 120-degree spin state, the peak of $\hat{S}(\bm q)$ at $\Gamma$ point rapidly decreases and is suppressed to almost zero. Subsequently, the peaks of $\hat{S}(\bm q)$ at K points gradually start to grow as seen in their simulated time profiles [Fig.~\ref{Fig03}]. In addition to $\hat{S}(\bm q)$, the vector spin chirality $C$ is also a good indicator of the 120-degree spin order. This quantity is defined by,
\begin{align}
\label{eq:Eq10}
C=\frac{1}{N}\sum_i |\bm C_i|,
\end{align}	
where $\bm C_i$ is a local spin chirality vector at the $i$th site defined with three localized spin vectors $\bm S_i$, $\bm S_{i+\hat{a}}$ and $\bm S_{i+\hat{a}+\hat{b}}$ located on three vertices of an up-pointing triangular plaquette as,
\begin{align}
\label{eq:Eq11}
\bm{C}_{i}=\frac{2}{3\sqrt{3}}
(\bm{S}_i\times\bm{S}_{i+\hat{a}}
+\bm{S}_{i+\hat{a}}\times\bm{S}_{i+\hat{a}+\hat{b}}
+\bm{S}_{i+\hat{a}+\hat{b}}\times\bm{S}_i).
\end{align}	
The simulated time profile of $C$ in Fig.~\ref{Fig03} shows that this quantity begins to grow immediately after the photoirradiation starts and eventually converge to almost unity when the 120-degree spin state emerges.

\begin{figure*}[tbh]
\centering
\includegraphics[scale=0.5]{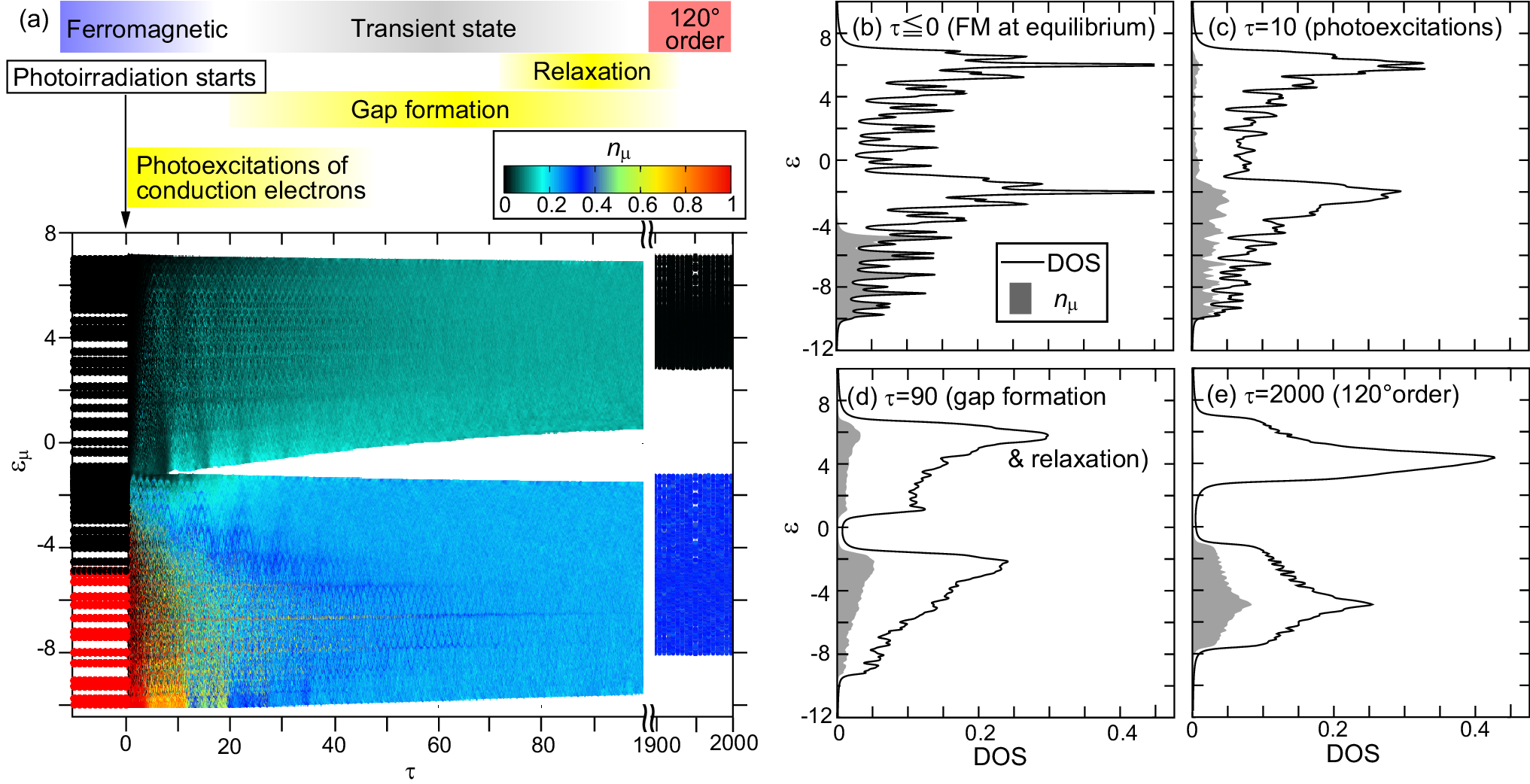}
\caption{(a) Simulated time evolutions of the eigenenergies $\varepsilon_\mu$ (dots) and the electron occupation of the corresponding eigenstates $n_\mu$ (colors) at early and final stages of the phase-transition process. (b)-(e) Simulated densities of states (DOS) and electron occupation at various stages of the phase-transition process, i.e., (b) before the photoirradiation ($\tau \leq 0$) where the system is in equilibrium with the ferromagnetic ground state, (c) immediately after starting the irradiation ($\tau=10$) where the photoexcitation of conduction electrons occur, (d) in the transient process ($\tau=90$) where the band gap is gradually formed, and (e) after sufficient duration ($\tau=2000$) where the system has the 120-degree spin order as a nonequilibrium steady state. The shaded areas indicate the electron-occupation rate of the bands. The parameters and conditions used for the simulations are the same as those for Fig.~\ref{Fig03}. This figure is taken and modified from Ref.~\cite{Inoue2022} {\copyright} 2022 American Physical Society.}
\label{Fig04}
\end{figure*}

The dynamical process and the physical mechanism of the photoinduced magnetic phase transition in the double-exchange model can be argued based on the simulated time-evolutions of electronic structures. Figure~\ref{Fig04}(a) shows the simulated time profiles of the electron band structures. More specifically, the dots represent eigenenergies $\varepsilon_\mu(\tau)$, while the colors represent the electron occupation $n_\mu(\tau)$ of the $\mu$th eigenstate at each moment. The eigenenergies $\varepsilon_\mu$ are obtained from the eigenequation,
\begin{align}
\label{eq:Eq12}
\mathcal{H}(\tau)\ket{\Psi_\mu(\tau)}=\varepsilon_\mu(\tau)\ket{\Psi_\mu(\tau)}.
\end{align}
Here $\mathcal{H}(\tau)$ is the time-dependent Hamiltonian, while $\ket{\Psi_\mu(\tau)}$ is the $\mu$th one-particle eigenstate where $\mu$ ($=1, 2,\ldots, 2N$) labels the eigenstates in ascending order with respect to the eigenenergies $\varepsilon_\mu$. On the other hand, the number of electrons that occupy the $\mu$th eigenstate $n_\mu(\tau)$ is calculated by,
\begin{align}
\label{eq:Eq13}
n_\mu(\tau)=\braket{\tilde{\Phi}(\tau)|\Psi_\mu(\tau)}\braket{\Psi_\mu(\tau)|\tilde{\Phi}(\tau)}.
\end{align}

The simulation results in Fig.~\ref{Fig04}(a) indicate that the dynamical process of the photoinduced phase transition is composed of three steps, that is, (Step 1) photoexcitation of electrons, (Step 2) formation of the band gap, and (Step 3) relaxation of the excited electrons. Through these steps, a special electronic structure called pseudo half-filling state is realized, to which the emergence of 120-degree spin state under the photoirradiation is attributable. We explain each of the three steps in detail below.\\

\noindent
{\bf (Step 1) Photoexcitation of electrons}\\
Before the light irradiation ($t<0$), the band structure is gapless, and each eigenstate below the Fermi level is occupied by a single electron as indicated by the red color in Fig.~\ref{Fig04}(a), while the states above the Fermi level are completely empty as indicated by the black color. This aspect can be clearly seen in the density of the states shown in Fig.~\ref{Fig04}(b). Right after the light irradiation starts, the electrons begin to be excited to higher-energy states within the continuum band. As the light irradiation starts at $t=0$, the colors in Fig.~\ref{Fig04}(a) indicate that the electrons are rapidly excited. A typical density of states at this stage is shown in Fig.~\ref{Fig04}(c).\\

\noindent
{\bf (Step 2) Formation of the band gap}\\
After a short period of light irradiation, the band gap begins to open, and its further growth separates the band into upper and lower parts. This gap formation originates from a bandwidth narrowing effect in the time-periodically driven systems, which is referred to as dynamical localization effect~\cite{Dunlap1986,Grossmann1991,Holthaus1992,Eckardt2005,Lignier2007,Kayanuma2008,Ishikawa2014,Ohmura2021}. In the photoirradiated tight-binding model, Peierls phases due to the time-dependent vector potential are added to the transfer integrals. At this time, the time-averaged transfer integrals are renormalized by the Bessel function coefficient $J_0(\mathcal{A}_{ij})$. As a result, the effective bandwidth is reduced in the photodriven systems~\cite{Yonemitsu2017,Kitayama2020,Kitayama2021a,Kitayama2021b,Kitayama2022}. When the bandwidth becomes smaller than the band splitting due to the Kondo exchange coupling, a gap is opened. The colors in Fig.~\ref{Fig04}(a) indicate that the band above the gap is partially occupied by electrons. A typical density of states at this stage is shown in Fig.~\ref{Fig04}(d).\\

\noindent
{\bf (Step 3) Relaxation of the excited electrons}\\
The photoexcitation of electrons continues to occur as long as the band gap is small even after the gap begins to open. When the gap grows sufficiently large, the excitation from the lower band to the upper band no longer occurs. Conversely, electrons excited to the upper band begin to fall to the lower band through losing their energy by dissipation. Note that in the numerical simulation, the effect of energy dissipation is taken into account by the Gilbert-damping term in the LLG equation. This means that the excited electrons lose their energy through coupling to the dissipative dynamics of localized spins via the Kondo coupling. As a result, after a long time, electrons in the upper band completely disappear as indicated by the uniform black color, while in the lower band, the excited electrons are uniformly distributed in all eigenstates constituting this band as indicated by the uniform blue color in Fig.~\ref{Fig04}(a). A typical density of states at this stage is shown in Fig.~\ref{Fig04}(e).\\

The electronic structure with the partially but uniformly occupied lower band as well as the empty upper band resembles the electronic structure in the half-filled system at equilibrium, in which the 120-degree spin order appears. It is known that the Fermi-surface nesting realized in the half-filled triangular double-exchange model at equilibrium favors the 120-degree spin order~\cite{Akagi2010,Akagi2012}. If all eigenstates constituting the lower band are partially but uniformly occupied by electrons, the electrons occupying the higher-energy part of the lower band are affected by similar nesting vectors in this quasi-half-filled situation. This situation is expected to stabilize the 120-degree spin order in the nonequilibrium system under light irradiation. It should be emphasized that the phenomenon that long-range magnetic ordered states are stabilized in photodriven double-exchange models through formation of the pseudo half-filling electron state can be expected not only in the triangular double-exchange model argued here but also in double-exchange models on various lattice structures in different dimensions. Indeed, we have confirmed that the same physics and phenomena appear universally in double-exchange models on cubic and Kagome lattices.

Finally, we would like to mention recent theoretical studies on the double-exchange models irradiated by microwave fields~\cite{Mochizuki2024,Eto2021,Eto2022,Eto2024a}. The double-exchange models describe composite systems of localized spins and conduction electrons coupled via exchange interactions, whose dynamics have different time scales. Specifically, the time scale of the dynamics of localized spins is of the order of picoseconds or nanoseconds with terahertz or gigahertz frequencies, which is relatively slow. On the other hand, the time scale of the dynamics of conduction electrons is of the order of femtoseconds, which is relatively fast. Because of these two different time scales in dynamics, the theoretical treatment differs between the case of irradiation with high-frequency electromagnetic waves (e.g., visible light) and the case of irradiation with low-frequency electromagnetic waves (e.g., microwaves). In the former case, as discussed in this section, the light electromagnetic field mainly interacts with the conduction electrons, whereas the localized spin cannot follow the high-frequency light. Therefore, the effect of light irradiation is considered through vector potentials attached to the transfer integrals. In the latter case, on the other hand, the microwave magnetic field can interact directly with the localized spins to induce their dynamics, whereas the conduction electrons feel the slowly time-varying fields from the localized spins as an adiabatic potential. Therefore, to investigate the effects of microwave irradiation, time evolutions of the localized spins induced by the microwave magnetic field are calculated using the LLG equation, and the states of conduction electrons are traced by treating the Hamiltonian with a given spin configuration at each moment within the adiabatic approximation. The double-exchange model has turned out to host rich topological magnetic states~\cite{Hayami2021R,Kawamura2025R}. Recent studies based on the above theoretical framework have predicted and proposed many interesting dynamical phenomena associated with real-space magnetic topology that can be expected when the double-exchange models hosting the topological magnetism are irradiated by microwave magnetic fields~\cite{Mochizuki2024,Eto2021,Eto2022,Eto2024a}.

To summarize this section, we have discussed a theoretical study on the photoinduced magnetic phase transition in the double-exchange model on a triangular lattice, which shows a transition from the ground-state ferromagnetic phase to the nonequilibrium 120-degree spin ordered phase under light irradiation. It has turned out that this phenomenon is caused by a specific electron-occupation state under irradiation named pseudo half-filling state, which is characterized by partial but uniform electron occupations of states constituting the low-energy band separated by the upper-energy band by a band gap. This electron-occupation state is realized by redistribution of photoexcited electrons in the low-energy band through relaxation after a band gap is formed by the exchange interactions and the bandwidth contraction due to the dynamical localization effect. The direct magnetic light-matter interactions between the light magnetic field and magnetization can hardly modulate the spatial configuration of localized spins because they are very weak. On the contrary, it is feasible to modulate the electronic structures such as band dispersions, band width, Fermi surfaces, and density of states with light electric field because the electric light-matter interactions between the light electric field and electron charges are significantly strong. In the double-exchange magnets, the spin-spin interactions are mediated by conduction electrons, and, thereby, their properties and the resulting long-range spin orders are governed by electronic structures. This aspect provides us with a precious opportunity for efficient photocontrol of magnetism via modulating the electronic structures by taking advantage of the strong electric light-matter interactions. The physical mechanism of the photoinduced phase transition associated with the pseudo half-filling state can be generally expected not only in the present triangular double-exchange model but also the double-exchange models with various lattice structures and spatial dimensions. 

The double-exchange model was originally applied to the colossal magnetoresistive manganite system~\cite{Tokura1999}. Recently, a new double-exchange material CeRh$_6$Ge$_4$ which realizes a ferromagnetic ground state on a triangular lattice has been discovered and studied intensively~\cite{Matsuoka2015,Kotegawa2019,BShen2020,WuY2021}. These materials might be suitable for experimental observations of the photoinduced magnetic phase transitions from ferromagnetic to antiferromagnetic phases predicted here. Moreover, novel materials which host rich topological magnetic phases described by the double-exchange models have been discovered and synthesized successively, e.g., MnSi$_{1-x}$Ge$_x$~\cite{Kanazawa2016,Fujishiro2019}, SrFeO$_3$~\cite{Ishiwata2020}, Gd$_2$PdSi$_3$~\cite{Kurumaji2019}, Gd$_3$Ru$_4$Al$_{12}$~\cite{Hirschberger2019}, EuPtSi~\cite{Kaneko2019}, and GdRu$_2$Si$_2$~\cite{Khanh2020}. We hope that research on the efficient photocontrol of magnetism in double-exchange magnets will be developed using this new class of materials.

\section{Highly efficient photoinduction of spin polarization in Rashba electron systems}
The inverse Faraday effect, i.e., the photoinduction of spin polarization with circularly polarized light, is one of the most important phenomena associated with magnetic responses of materials irradiated by light~\cite{Mochizuki2018,Tanaka2020,Pitaevskii1961,Pershan1966,Ziel1965,Hertel2006,Battiato2014,Takayoshi2014a,Takayoshi2014b,Sato2016,Stanciu2007b}. This effect has been studied intensively so far both experimentally and theoretically. From the theoretical point of view, Pitaevskii proposed a phenomenological theory for continuum media as pioneering work on this effect in 1961~\cite{Pitaevskii1961}. In the same year, Pershan and coworkers discussed this phenomenon for an isolated atom with spin-orbit interactions using a perturbation theory with respect to the light electric field $\bm E$~\cite{Pershan1966}. They found that an effective magnetic field and net magnetization perpendicular to the circular-polarization plane, magnitudes of which are proportional to $E^2$, are generated when this system is irradiated by circularly polarized light. Subsequently, Hertel studied this phenomenon in metals by treating the electrons as a collisionless plasma in 2006~\cite{Hertel2006}. He found that the induced spin polarization is proportional to the square of light amplitude $E$ and inversely proportional to the cube of light frequency $\omega$, i.e., $M \propto E^2/\omega^3$. Recently, Mochizuki and coworkers studied this phenomenon in an electron system with the Rashba spin-orbit interaction by numerically simulating the spatiotemporal dynamics of a Gaussian wave packet under light irradiation in 2018~\cite{Mochizuki2018}. They found that highly efficient photoinduction of spin polarization is possible in the Rashba electron system. In these previous studies, the inverse Faraday effects are discussed only at phenomenological levels. Hence, the effects of band dispersions, Fermi surfaces, and electron filling on this phenomenon have not been clarified sufficiently.

In 2020, an elaborate theory of the inverse Faraday effect based on a microscopic tight-binding model was proposed by Tanaka and coworkers~\cite{Tanaka2020}, in which effects of the band formation as well as those of the spin-orbit interaction and circularly polarized light are considered. They performed both unbiased numerical simulations of the magnetization dynamics and constructed a modern theory based on the Floquet theorem. The above-mentioned light-parameter dependence of the induced spin polarization, i.e., $M \propto E^2/\omega^3$ was reproduced by this microscopic model as well. Moreover, this study uncovered sensitive dependence of the sign and magnitude of the light-induced spin polarization on the size of the Fermi surfaces or the electron filling. This dependence never appears in previous phenomenological theories and can be described only by microscopic models for solids with electron band structures. The quantitative discussion based on microscopic theory provides important insights into the photoinduced spin polarization in real materials.

\begin{figure}[tb]
\centering
\includegraphics[scale=0.5]{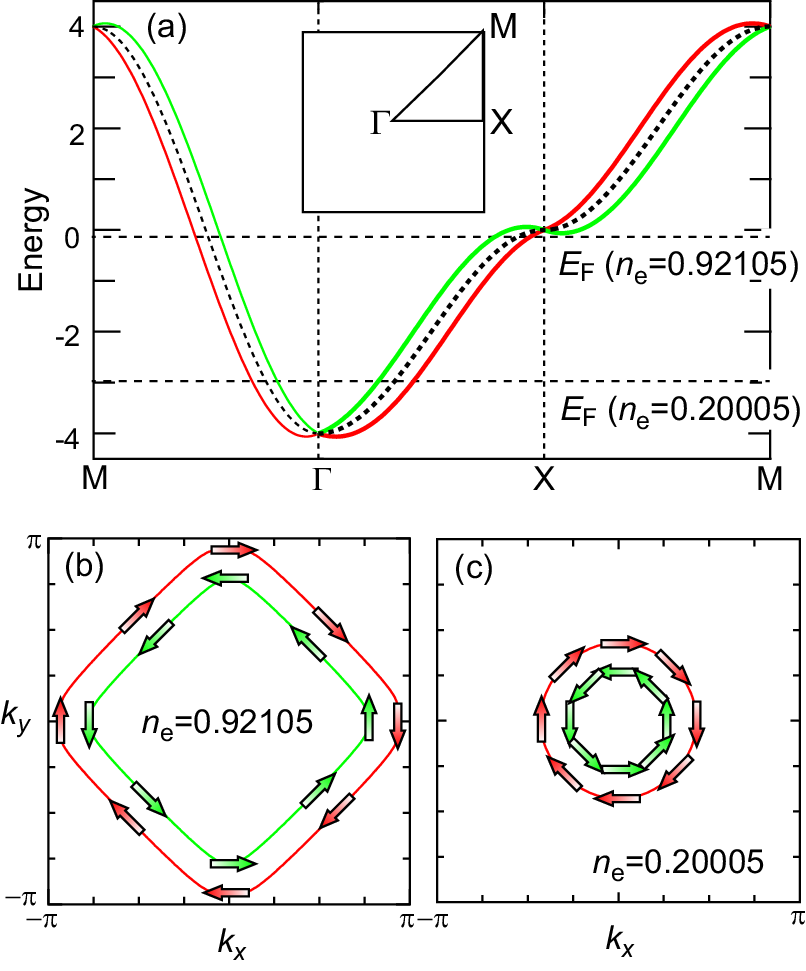}
\caption{(a) Band dispersion relations of the tight-binding model with the Rashba spin-orbit interaction on a square lattice for the Rashba parameter of $\alpha_{\rm R}=0.5$. Those without the Rashba spin-orbit interaction are also shown by a dashed line. Fermi levels are indicated by horizontal dashed lines for higher electron filling of $n_{\rm e}$=0.92105 and lower electron filling of $n_{\rm e}$=0.20005. (c),(d) Fermi surfaces for $n_{\rm e}$=0.92105 (c) and $n_{\rm e}$=0.20005 (d). The arrows indicate the spin orientations on the Fermi surfaces. This figure is taken and modified from Ref.~\cite{Tanaka2020} {\copyright} 2020 IOP Publishing.}
\label{Fig05}
\end{figure}
In Ref.~\cite{Tanaka2020}, the tight-binding model on a square lattice is employed to describe the two-dimensional Rashba electron system at equilibrium. The Hamiltonian is given by,
\begin{align}
\label{eq:Eq14}
\mathcal{H}=\mathcal{H}_{\rm kin}+\mathcal{H}_{\rm so},
\end{align}
where
\begin{align}
\label{eq:Eq15}
\mathcal{H}_{\rm kin}=\sum_{\bm k,\sigma}\varepsilon_{\bm k} c^{\dagger}_{\bm k\sigma}c_{\bm k\sigma},
\end{align}
with
\begin{align}
\label{eq:Eq16}
\varepsilon_{\bm k}=2t \left\{\cos k_x + \cos k_y \right\},
\end{align}
and 
\begin{align}
\label{eq:Eq17}
\mathcal{H}_{\rm so}=-\alpha_{\rm R} \sum_{\bm k}\left\{
\sin(k_x)c^{\dagger}_{\bm k\sigma} \sigma_y^{\sigma\sigma'} c_{\bm k\sigma'} -\sin(k_y)c^{\dagger}_{\bm k\sigma} \sigma_x^{\sigma\sigma'} c_{\bm k\sigma'}\right\}.
\end{align}
Here the term $\mathcal{H}_{\rm kin}$ describes kinetic energies of electrons with the nearest-neighbor transfer integrals $t$, while the term $\mathcal{H}_{\rm so}$ describes the Rashba spin-orbit interaction where $\alpha_{\rm R}$ is the coupling constant called Rashba parameter~\cite{Rashba1960,WinklerBook2003,Shimizu2020}. Figure~\ref{Fig05}(a) shows the band dispersion relation with a splitting due to the Rashba spin-orbit interaction when $t$=1 and $\alpha_{\rm R}/t$=0.5. Note that an unphysically large Rashba parameter is adopted to exaggerate the Rashba band splitting here, but a realistic value was used for the actual numerical simulations. Figures~\ref{Fig05}(b) and (c) show the Fermi surfaces for higher electron filing of $n_{\rm e}$=0.92105 and lower electron filing of $n_{\rm e}$=0.20005, respectively, which are also split by the Rashba spin-orbit interaction.

Here, a situation where the system is irradiated by circularly polarized light is considered. The ac electric field $\bm E(\tau)$ and ac magnetic field $\bm B(\tau)$ of the light are given by,
\begin{align}
\label{eq:Eq18}
&\bm E(\tau)=E_0\beta(\tau) \left(\cos\omega \tau, \sin\omega \tau \right),
\\
\label{eq:Eq19}
&\bm B(\tau)=B_0\beta(\tau) \left(\sin\omega \tau, -\cos\omega \tau \right),
\end{align}
where $\beta(\tau)=1-e^{-\tau^2/\tau^2_d}$ is a prefactor introduced to mimic a gradual increase of the light intensity. Here we adopt the natural units $e$=$\hbar$=$c$=1, and the relation $E_0/B_0=c$ holds where $c$ is the speed of light. The vector potential for these light electromagnetic fields is given by,
\begin{align}
\label{eq:Eq20}
{\bm A}(\tau)=-\int^{\tau}_{0}{\bm E}(\tau')d\tau'.
\end{align}
Using the vector potential $A(\tau)=(A_x(\tau),A_y(\tau),A_z(\tau))$, the time-dependent Hamiltonian for the photodriven Rashba electron system is given by,
\begin{align}
\label{eq:Eq21}
{\mathcal H}(\tau)
={\mathcal H}_{\rm kin}(\tau)+{\mathcal H}_{\rm so}(\tau)+{\mathcal H}_{\rm Zeeman}(\tau).
\end{align}
The kinetic-energy term ${\mathcal H}_{\rm kin}(\tau)$ is give by,
\begin{align}
\label{eq:Eq22}
{\mathcal H}_{\rm kin}(\tau)=\sum_{<i,j>}t \exp\left[-i\bm A(\tau) \cdot \bm e_{ij} \right] c^\dagger_i c_j.
\end{align}
After the Fourier transformation, its momentum representation is obtained as,
\begin{align}
\label{eq:Eq23}
{\mathcal H}_{\rm kin}(\tau)=\sum_{\bm k,\sigma}\varepsilon_{\bm k, A}(\tau)\;
c^\dagger_{\bm k\sigma}c_{\bm k\sigma},
\end{align}
with
\begin{align}
\label{eq:Eq24}
\varepsilon_{\bm k,A}(\tau)=-2t \left\{\cos(k_x+A_x(\tau))+\cos(k_y+A_y(\tau)) \right\}.
\end{align}
The Rashba spin-orbit interaction term ${\mathcal H}_{\rm so}(\tau)$ and the Zeeman-coupling term ${\mathcal H}_{\rm Zeeman}(\tau)$ are given by,
\begin{align}
\label{eq:Eq25}
{\mathcal H}_{\rm so}(\tau)
&=-\alpha_{\rm R}\sum_{\bm k}\left\{
\sin(k_x+A_x(\tau))c^\dagger_{\bm k\sigma} \sigma_y^{\sigma\sigma'} c_{\bm k\sigma'} \right.
\notag \\ &\left. \hspace{1.5cm}
-\sin(k_y+A_y(\tau))c^\dagger_{\bm k\sigma} \sigma_x^{\sigma\sigma'} c_{\bm k\sigma'}
\right\}.
\end{align}
and
\begin{align}
\label{eq:Eq26}
{\mathcal H}_{\rm Zeeman}(\tau)=
\sum_{\bm k}\left\{
 B_x(\tau) c^\dagger_{\bm k\sigma} \sigma_x^{\sigma\sigma'} c_{\bm k\sigma'}
+B_y(\tau) c^\dagger_{\bm k\sigma} \sigma_y^{\sigma\sigma'} c_{\bm k\sigma'}
\right\}
\end{align}

The matrix-representation of the time-dependent Hamiltonian is given in the form,
\begin{align}
\label{eq:Eq27}
{\mathcal H}(\tau)=\sum_{\bm k}(c^\dagger_{\bm k\uparrow},c^\dagger_{\bm k\downarrow})
\hat{\mathcal H}_{\bm k}(\tau)
\begin{pmatrix}
c_{\bm k\uparrow}\\
c_{\bm k\downarrow}
\end{pmatrix},
\end{align}
where
\begin{align}
\label{eq:Eq28}
\hat{\mathcal H}_{\bm k} \equiv
\begin{pmatrix}
\varepsilon_{\bm k,A}(\tau) & \gamma_{\bm k,A}(\tau)\;+\;b(\tau) \\
\gamma^{\ast}_{\bm k,A}(\tau)\;+\;b^{\ast}(\tau) & \varepsilon_{\bm k,A}(\tau)
\end{pmatrix},
\end{align}
with
\begin{align}
\label{eq:Eq29}
&\gamma_{\bm k,A}(\tau)=\alpha_{\rm R}\left\{ i\sin(k_x+A_x(\tau))+\sin(k_y+A_y(\tau)) \right\},
\\
\label{eq:Eq30}
&b(\tau)=B_0\beta(\tau) \left\{ \sin\omega\tau+i\cos\omega\tau \right\}.
\end{align}

We simulate the photoinduced dynamics of electrons by numerically solving the time-dependent Schr\"odinger equation,
\begin{align}
\label{eq:Eq31}
i\frac{\partial}{\partial\tau}|\Psi_{\bm k,\nu}(\tau)\rangle ={\mathcal H}(\tau)|
\Psi_{\bm k,\nu}(\tau)\rangle
\end{align}
where $|\Psi_{\bm k,\nu}(\tau)\rangle$ is the $\nu$th one-particle wavefunction ($\nu$=1,2) at wave vector $\bm k$. This equation can be solved formally as~\cite{Terai1993,Kuwabara1995},
\begin{align}
\label{eq:Eq32}
|\Psi_{\bm k,\nu}(\tau+\Delta \tau)\rangle = \exp
\left[-i\Delta \tau {\mathcal H}_{\bm k}(\tau+\Delta \tau/2)\right]
|\Psi_{\bm k,\nu}(\tau)\rangle
\end{align}
with $\Delta\tau$=0.01. The numerical simulations are performed using a system with $N=200 \times 200$ sites for the Rashba parameter $\alpha_{\rm R}$=0.1, light frequency $\omega$=1, the light amplitude $E_0$=0.1, and $\tau_{\rm d}$=50. These parameter values correspond to $\alpha_{\rm R}$=0.1 eV, $\omega/2\pi$=242 THz, $E_0$=2 MV/cm, and $\tau_{\rm d}$=33 fs, respectively, when we assume $t$=1 eV and $a$=5\AA as the units of energy and length.

\begin{figure}[tb]
\includegraphics[scale=0.5]{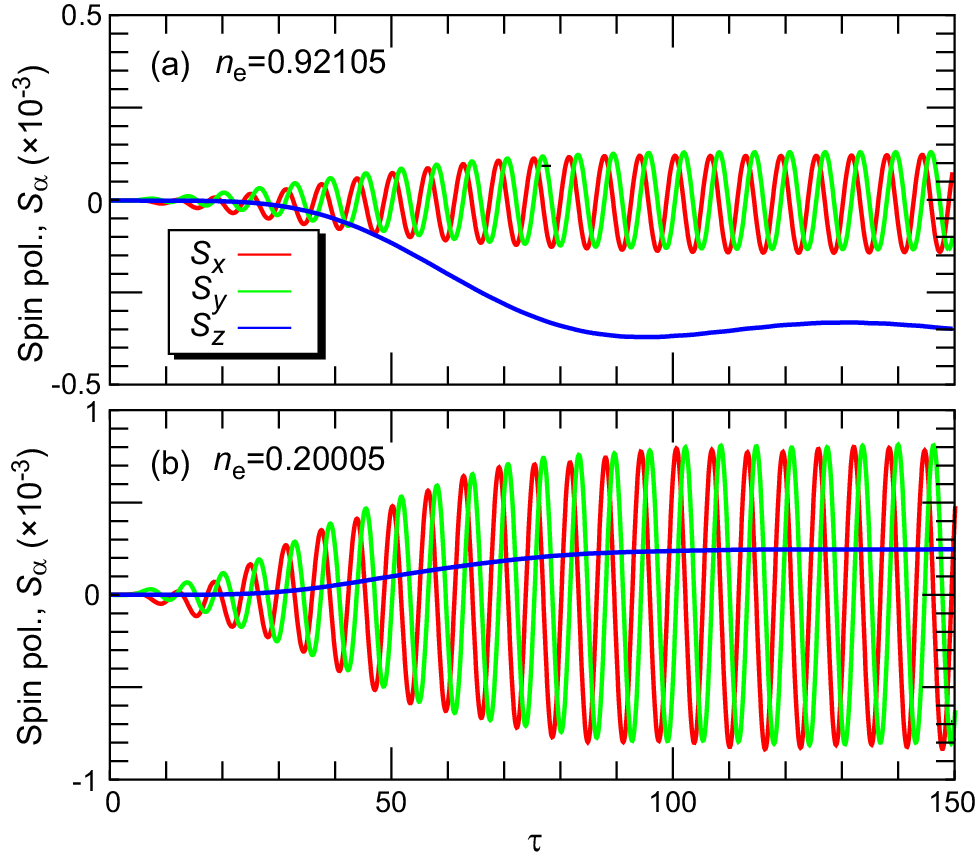}
\caption{Simulated time profiles of the induced ne spin polarization $\bm S(\tau)=(S_x(\tau), S_y(\tau), S_z(\tau))$ for (a) higher electron filling of $n_{\rm e}$=0.92105 and (b) lower electron filling of $n_{\rm e}$=0.20005. The simulations are performed for $\alpha_{\rm R}$=0.1, $\omega$=1, $E_0$=0.1, and $\tau_d$=50. The unit conversions are given in Table \ref{tab:unitconv}. This figure is taken and modified from Ref.~\cite{Tanaka2020} {\copyright} 2020 IOP Publishing.}
\label{Fig06}
\end{figure}
Figures~\ref{Fig06}(a) and (b) show the simulated time profiles of $x$-, $y$- and $z$-axis components of the spin polarization $S_{\alpha}$ ($\alpha$=$x$,$y$,$z$) for higher electron filling of $n_{\rm e}$=0.92105 and lower electron filling of $n_{\rm e}$=0.20005, respectively. The spin polarization $S_{\alpha}$ is defined as,
\begin{align}
\label{eq:Eq33}
S_{\alpha}=\frac{1}{2N_{\bm k}}\sum_{\bm k}\langle c^\dagger_{\bm k\sigma}
\sigma_{\alpha}^{\sigma\sigma'} c_{\bm k\sigma'}\rangle
\end{align}
where $N_k$ is the number of $\bm k$ points between the two Fermi surfaces. In these figures, we find that the in-plane components $(S_x,S_y)$ oscillating around zero. One the contrary, the out-of-plane components $S_z$ are saturated to a finite value, which indicates induction of the spin polarization by photoirradiation. Importantly, the sign of $S_z$ depends on the electron filling Specifically, the induced spin polarization is negative ($S_z<0$) for the higher electron filling, while it is positive ($S_z>0$) for the lower electron filling. 

\begin{figure}
\centering
\includegraphics[scale=0.5]{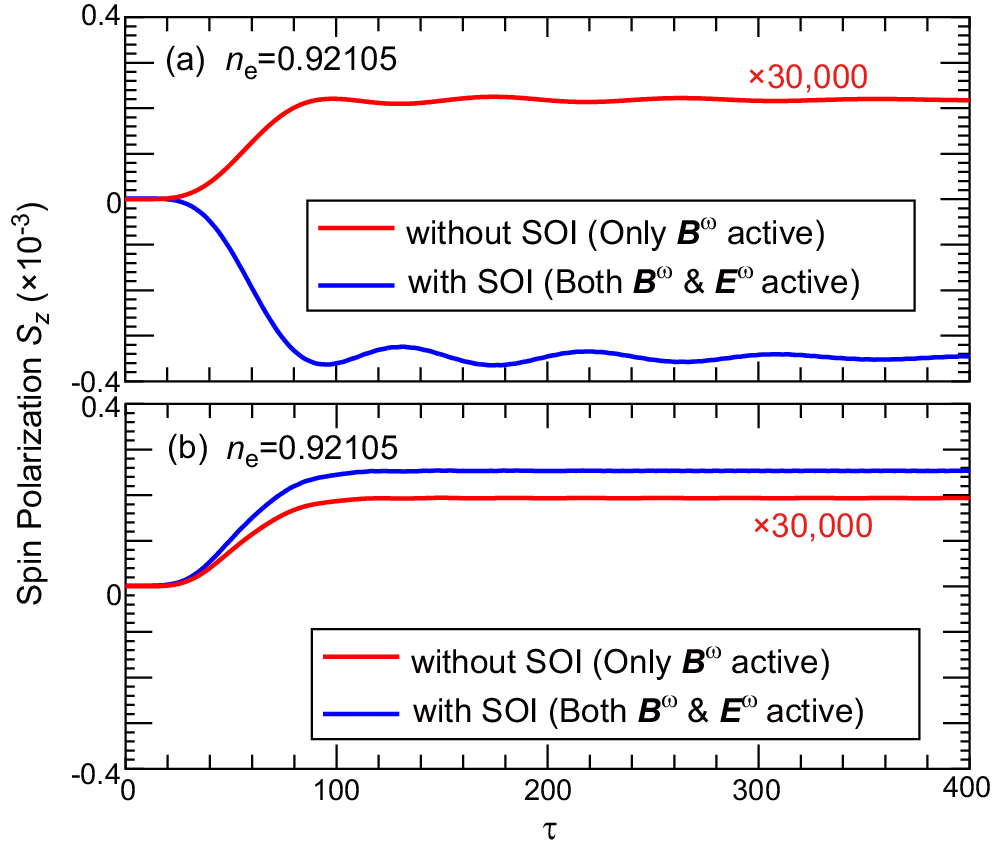}
\caption{(a) Simulated time profiles of the photoinduced spin polarization $S_z(\tau)$ in the presence ($\alpha_{\rm R}=0.1$) and absence ($\alpha_{\rm R}=0$) of the Rashba spin-orbit interaction for the higher electron filling of $n_{\rm e}$=0.92105. (b) Those for the lower electron filling of $n_{\rm e}$=0.20005. When the Rashba spin-orbit interaction is present, the spin polarization is induced both by the magnetic-field ($\bm B^\omega$) and electric-field ($\bm E^\omega$) components of light, whereas it is induced only by the $\bm B^\omega$ component when the Rashba spin-orbit interaction is absent. The simulations are performed for $\alpha_{\rm R}=0.1$, $\omega=1$, $E_0=0.1$, and $\tau_d=50$. The unit conversions are given in Table~\ref{tab:unitconv}. This figure is taken and modified from Ref.~\cite{Tanaka2020} {\copyright} 2020 IOP Publishing.}
\label{Fig07}
\end{figure}
Figure~\ref{Fig07}(a) shows comparison between the spin-polarizations $S_z$ in the presence and absence of the spin-orbit interaction for the higher electron filling of $n_{\rm e}$=0.92105, while Fig.~\ref{Fig07}(b) shows those for the lower electron filling of $n_{\rm e}$=0.20005. Here we note that when the spin-orbit interaction is absent, only the light magnetic field $\bm B^\omega$ contributes to the spin-polarization induction. On the contrary, both the light magnetic field $\bm B^\omega$ and electric field $\bm E^\omega$ contribute to the spin-polarization induction in the presence of the spin-orbit interaction. These figures indicate,
\begin{itemize}
\item The $\bm B^\omega$-induced spin polarization is four orders of magnitude smaller than the $\bm E^\omega$-induced spin polarization.\\
\item The $\bm B^\omega$-induced spin polarization is always positive for the left-handed circularly polarized light considered in the present simulations irrespective of the electron filling.\\
\item The filling-dependence of the sign of induced spin polarization governed by the $\bm E^\omega$-induced component.
\end{itemize}

As argued above, the numerical simulations based on the tight-binding model for the Rashba electron system revealed the induction of static spin polarization perpendicular to the light-polarization plane under irradiation with circularly polarized light. It was demonstrated that the light electric field can contribute to the spin-polarization induction in the presence of spin-orbit interaction, and its contribution is overwhelmingly larger than the contribution from light magnetic field. Therefore, highly efficient photoinduction of spin polarization is possible in the Rashba electron system. Moreover, the sign of the photoinduced spin polarization turns out to vary depending on the electron filling in the Rashba electron systems, which originates from the component induced by light electric field. Moreover, the time-averaged spin polarization turns out to be proportional to $E_0^2/\omega^3$ where $E_0$ and $\omega$ are the amplitude and frequency (angular velocity) of light, respectively (not shown).

To understand the above findings in the numerical simulations, an analytical theory based on the Floquet theorem was constructed. The Floquet theorem is a temporal version of the Bloch theorem, with which a problem of system driven by a time-periodic external field can be mapped onto an effective static eigenvalue problem with a matrix of infinite dimensions. The theory is formulated as follows. We start with the Schr\"odinger equation for a time-periodically driven system, which is given by,
\begin{align}
\label{eq:Eq34}
i\frac{\partial}{\partial\tau}|\Psi(\tau)\rangle=\mathcal{H}(\tau)|\Psi(\tau)\rangle,
\end{align}
where $\mathcal{H}(\tau)=\mathcal{H}(\tau+T)$ is the time-periodic Hamiltonian with a time period of $T=2\pi/\omega$. According the Floquet theorem, the wavefunction of this system can be written as,
\begin{align}
\label{eq:Eq35}
|\Psi(\tau)\rangle=e^{-i\epsilon \tau}|\Phi(\tau)\rangle,
\end{align}
with
\begin{align}
\label{eq:Eq36}
|\Phi(\tau)\rangle=|\Phi(\tau+T)\rangle.
\end{align}
Because the Hamiltonian $\mathcal{H}(\tau)$ and the Floquet states $|\Phi(\tau)\rangle$ are time-periodic, we introduce the Fourier transforms as,
\begin{align}
\label{eq:Eq37}
&\mathcal{H}(\tau)=\sum_m e^{-im\omega\tau}\mathcal{H}_m,
\\
\label{eq:Eq38}
&|\Phi(\tau)\rangle=\sum_m e^{-im\omega\tau}|\Phi_m\rangle,
\end{align}
with
\begin{align}
\label{eq:Eq39}
&\mathcal{H}_m=\frac{1}{T}\int^{T}_{0}\mathcal{H}(\tau)e^{im\omega\tau}d\tau,
\\
\label{eq:Eq40}
&|\Phi_m\rangle=\frac{1}{T}\int^{T}_{0}|\Phi(\tau)\rangle e^{im\omega\tau}d\tau.
\end{align}
Using thus defined Fourier components, the time-dependent Schr\"odinger equation can be rewritten as,
\begin{align}
\label{eq:Eq41}
\sum_n \left( \mathcal{H}_{m-n}-m\omega\delta_{m,n} \right)
|\Phi_n\rangle =\varepsilon|\Phi_m\rangle,
\end{align}
where $\varepsilon$ is the Floquet quasienergy, and $m$ and $n$ are the Floquet indices that represent the number of photons. This expression no longer contains time, and thus the problem is reduced to the static eigenvalue problem. This equation is written in the matrix representation as,
\begin{align}
\label{eq:Eq42}
\begin{pmatrix}
\ddots & \ddots & & & \\
\ddots & \mathcal{H}_0-\omega & \mathcal{H}_1 & \mathcal{H}_2 & \\
& \mathcal{H}_{-1} & \mathcal{H}_0 & \mathcal{H}_1 & \\
& \mathcal{H}_{-2} & \mathcal{H}_{-1} & \mathcal{H}_0+\omega & \ddots \\
&  &  & \ddots & \ddots
\end{pmatrix}
\begin{pmatrix}
\vdots \\
|\Phi_1\rangle \\
|\Phi_0\rangle \\
|\Phi_{-1}\rangle \\
\vdots
\end{pmatrix}
=\varepsilon
\begin{pmatrix}
\vdots \\
|\Phi_1\rangle \\
|\Phi_0\rangle \\
|\Phi_{-1}\rangle \\
\vdots
\end{pmatrix}
\end{align}
In the high-frequency limit of $\omega \gg t$, an effective Hamiltonian ${\mathcal H}_{\rm eff}$ in the zero-photon subspace can be derived by the high-frequency expansion as~\cite{Kitagawa2011,Lindner2011,Eckardt2015},
\begin{align}
\label{eq:Eq43}
{\mathcal H}_{\rm eff}={\mathcal H}_{0}-\frac{1}{\omega}[
{\mathcal H}_{1},{\mathcal H}_{-1}]
\end{align}
where
\begin{align}
\label{eq:Eq44}
\mathcal{H}_m=\frac{1}{T}\int^{T}_{0}\mathcal{H}(\tau)e^{im\omega\tau}d\tau
\hspace{0.5cm} (m=0,\pm 1).
\end{align}
The time-dependent Hamiltonian ${\mathcal H}(\tau)$ for the photodriven system is given in the momentum-representation form as,
\begin{align}
\label{eq:Eq45}
{\mathcal H}(\tau)=\sum_k(c^\dagger_{\bm k\uparrow},c^\dagger_{\bm k\downarrow})
\begin{pmatrix}
\varepsilon_{\bm k,A}(\tau) & \gamma_{\bm k,A}(\tau) \\
\gamma^{\ast}_{\bm k,A}(\tau) & \varepsilon_{\bm k,A}(\tau)
\end{pmatrix}
\begin{pmatrix}
c_{\bm k\uparrow}\\
c_{\bm k\downarrow}
\end{pmatrix},
\end{align}
where
\begin{align}
\label{eq:Eq46}
&\varepsilon_{\bm k,A}(\tau)=-2t \left\{\cos(k_x+A_x(\tau))+\cos(k_y+A_y(\tau))\right\},
\\
\label{eq:Eq47}
&\gamma_{\bm k,A}(\tau)=\alpha_{\rm R} \left\{i\sin(k_x+A_x(\tau))+\sin(k_y+A_y(\tau))\right\},
\end{align}
with
\begin{align}
\label{eq:Eq48}
{\bm A}(\tau)&\equiv (A_x(\tau), A_y(\tau)) \notag \\
&=A_0 (\cos\omega \tau, \sin\omega \tau) \notag \\
&=\frac{E_0}{\omega} (\cos\omega \tau, \sin\omega \tau).
\end{align}
After some algebra, we obtain the following expression for ${\mathcal H}_{0}$,
\begin{align}
\label{eq:Eq49}
{\mathcal H}_0
=\sum_k(c^\dagger_{\bm k\uparrow},c^\dagger_{\bm k\downarrow})
\begin{pmatrix}
\tilde \varepsilon_{\bm k,A} & \tilde \gamma_{\bm k,A} \\
\tilde \gamma^{\ast}_{\bm k,A} & \tilde \varepsilon_{\bm k,A}
\end{pmatrix}
\begin{pmatrix}
c_{\bm k\uparrow}\\
c_{\bm k\downarrow}
\end{pmatrix}
\end{align}
where
\begin{align}
\label{eq:Eq50}
&\tilde \varepsilon_{\bm k,A}
=\left(1-\frac{A_0^2}{4}\right)\varepsilon_{\bm k},
\quad
\tilde \gamma_{\bm k,A}
=\left(1-\frac{A_0^2}{4}\right)\gamma_{\bm k}
\\
\label{eq:Eq51}
&\gamma_{\bm k} \equiv \alpha_{\rm R} \left\{i\sin(k_x)+\sin(k_y)\right\}.
\end{align}
On the other hand, we obtain the following expressions for ${\mathcal H}_{1}$ and ${\mathcal H}_{-1}$,
\begin{align}
\label{eq:Eq52}
{\mathcal H}_{1}
=\sum_{\bm k,\sigma}s({\bm k})c^{\dagger}_{\bm k\sigma}c_{\bm k\sigma}
+\sum_{\bm k} \left\{t({\bm k}) c^{\dagger}_{\bm k\downarrow}c_{\bm k\uparrow}
+u({\bm k}) c^{\dagger}_{\bm k\uparrow}c_{\bm k\downarrow}\right\}
\end{align}
and
\begin{align}
\label{eq:Eq53}
{\mathcal H}_{-1}
=\sum_{\bm k,\sigma}s^{\ast}({\bm k})c^{\dagger}_{\bm k\sigma}c_{\bm k\sigma}
+\sum_{\bm k} \left\{u({\bm k}) c^{\dagger}_{\bm k\downarrow}c_{\bm k\uparrow}
+t({\bm k}) c^{\dagger}_{\bm k\uparrow}c_{\bm k\downarrow}\right\}
\end{align}
where
\begin{align}
\label{eq:Eq54}
&s({\bm k})=-tA_0 \left\{i\sin k_x + \sin k_y \right\},
\\
\label{eq:Eq55}
&t({\bm k})=\frac{\alpha_{\rm R}A_0}{2}\left\{\cos k_x + \cos k_y \right\},
\\
\label{eq:Eq56}
&u({\bm k})=\frac{\alpha_{\rm R}A_0}{2}\left\{-\cos k_x + \cos k_y \right\}.
\end{align}
We further obtain,
\begin{align}
\label{eq:Eq57}
-\frac{[{\mathcal H}_{1},{\mathcal H}_{-1}]}{\omega}
&=\sum_{\bm k}\left\{u({\bm k})^2-t({\bm k})^2\right\}
\left(c^{\dagger}_{\bm k\uparrow}c_{\bm k\uparrow}-c^{\dagger}_{\bm k\downarrow}c_{\bm k\downarrow}\right)
\notag \\
&=\frac{\alpha_{\rm R}^2  A_0^2}{\omega}\sum_{\bm k}\cos k_x \cos k_y
\left(c^{\dagger}_{\bm k\uparrow}c_{\bm k\uparrow}-c^{\dagger}_{\bm k\downarrow}c_{\bm k\downarrow}\right)
\notag \\
&=\frac{\alpha_{\rm R}^2 E_0^2}{\omega^3}\sum_{\bm k}\cos k_x \cos k_y
\left(c^{\dagger}_{\bm k\uparrow}c_{\bm k\uparrow}-c^{\dagger}_{\bm k\downarrow}c_{\bm k\downarrow}\right).
\end{align}
Eventually, we obtain the effective Floquet Hamiltonian in the high-frequency limit as,
\begin{align}
\label{eq:Eq58}
{\mathcal H}_{\rm eff}
&={\mathcal H}_{0}-\frac{1}{\omega}[{\mathcal H}_{1},{\mathcal H}_{-1}]
\notag \\
&=\sum_{\bm k}(c^\dagger_{\bm k\uparrow},c^\dagger_{\bm k\downarrow})
\begin{pmatrix}
\tilde{\varepsilon}_{\bm k,A}+h^{\rm eff}_{\bm k} & \tilde{\gamma}_{\bm k,A} \\
\tilde{\gamma}^{\ast}_{\bm k,A} & \tilde{\varepsilon}_{\bm k,A}-h^{\rm eff}_{\bm k}
\end{pmatrix}
\begin{pmatrix}
c_{\bm k\uparrow}\\
c_{\bm k\downarrow}
\end{pmatrix},
\end{align}
where
\begin{align}
\label{eq:Eq59}
&\tilde{\varepsilon}_{\bm k,A}=\left(1-\frac{E^2_0}{4\omega^2}\right)\varepsilon_{\bm k}, 
\quad\quad
\tilde{\gamma}_{\bm k,A}=\left(1-\frac{E^2_0}{4\omega^2}\right)\gamma_{\bm k}.
\end{align}
Noticeably, the quantity $h^{\rm eff}_{\bm k}$ in the effective Floquet Hamiltonian ${\mathcal H}_{\rm eff}$ can be regarded as an effective magnetic field in the $z$ direction. Importantly, this effective magnetic field is dependent on the momentum ${\bm k}$ as
\begin{align}
\label{eq:Eq60}
h^{\rm eff}_{\bm k}=-\frac{\alpha_{\rm R}^2 E_0^2}{\omega^3}\cos k_x \cos k_y.
\end{align}
This expression indicates that the sign of the effective magnetic field is governed by the momentum $\bm k$.

The contribution from each momentum point to the spin polarization $s^z_{\bm k}$ can also be evaluated using the formula,
\begin{align}
\label{eq:Eq61}
s^z_{\bm k}=\frac{1}{2}\left(
 \langle c^\dagger_{\bm k\uparrow}c_{\bm k\uparrow}\rangle
-\langle c^\dagger_{\bm k\downarrow}c_{\bm k\downarrow}\rangle
\right).
\end{align}
The concrete expression of $s^z_{\bm k}$ is calculated by diagonalizing the Hamiltonian ${\mathcal H}_{\rm eff}$. Because the Hamiltonian is a $2 \times 2$ matrix, two eigenenergies $\widetilde{E}_{-}(\bm k)$ and $\widetilde{E}_{+}(\bm k)$ ($\widetilde{E}_{-}(\bm k)<\widetilde{E}_{+}(\bm k)$) are obtained at each momentum point $\bm k$, which correspond to the band splitting due to the Rashba spin-orbit interaction, i.e., the Rashba splitting. The eigenvectors for respective eigenenergies are $(u_{-}(\bm k),v_{-}(\bm k))$ and $(u_{+}(\bm k),v_{+}(\bm k))$, where
\begin{align}
\label{eq:Eq62}
&u_{\pm}(\bm k)=\frac{1}{\sqrt{2}}\frac{\tilde{\gamma}_{\bm k,A}}{|\tilde{\gamma}_{\bm k,A}|}
\left(
1\pm \frac{h^{\rm eff}_{\bm k}}{\sqrt{(h^{\rm eff}_{\bm k})^2+|\tilde{\gamma}_{\bm k,A}|^2}}
\right)^{1/2},
\\
\label{eq:Eq63}
&v_{\pm}(\bm k)=\pm\frac{1}{\sqrt{2}}
\left(
1\mp \frac{h^{\rm eff}_{\bm k}}{\sqrt{(h^{\rm eff}_{\bm k})^2+|\tilde{\gamma}_{\bm k,A}|^2}}
\right)^{1/2}.
\end{align}
Using these expressions, the contribution $s^z_{\bm k}$ is calculated as,
\begin{align}
\label{eq:Eq64}
s^z_{\bm k}&=\frac{1}{2}
\left\{
\left(\left|u_{-}(\bm k)\right|^2-\left|v_{-}(\bm k)\right|^2\right)f(\widetilde{E}_{-}\left(\bm k)\right)
\right.
\notag \\ &\hspace{1.0cm}
\left.+
\left(\left|u_{+}(\bm k)\right|^2-\left|v_{+}(\bm k)\right|^2\right)f(\widetilde{E}_{+}\left(\bm k)\right)
\right\},
\end{align}
where $f(E)$ is the Fermi distribution function at zero temperature, which is give by,
\begin{align}
\label{eq:Eq65}
f(\widetilde{E}_{\pm}(\bm k)) \equiv 1
-\theta(\widetilde{E}_{\pm}(\bm k)-\widetilde{\varepsilon}_F).
\end{align}
Here $\widetilde{\varepsilon}_F$ is the Fermi energy for ${\mathcal H}_{\rm eff}$ and $\theta(x)$ is the Heviside step function. Eventually, we obtain,
\begin{align}
\label{eq:Eq66}
s^z_{\bm k}=\left\{
\begin{array}{cl}
0 & f(\widetilde{E}_{-}(\bm k))=f(\widetilde{E}_{+}(\bm k))=1\\
0 & f(\widetilde{E}_{-}(\bm k))=f(\widetilde{E}_{+}(\bm k))=0\\
{\displaystyle 
-\frac{1}{2}\frac{h^{\rm eff}_{\bm k}}{\sqrt{(h^{\rm eff}_{\bm k})^2+|\tilde{\gamma}_{\bm k,A}|^2}}
}
& f(\widetilde{E}_{-}(\bm k))=1, f(\widetilde{E}_{+}(\bm k))=0.
\end{array}
\right.
\end{align}

\begin{figure}
\centering
\includegraphics[scale=0.5]{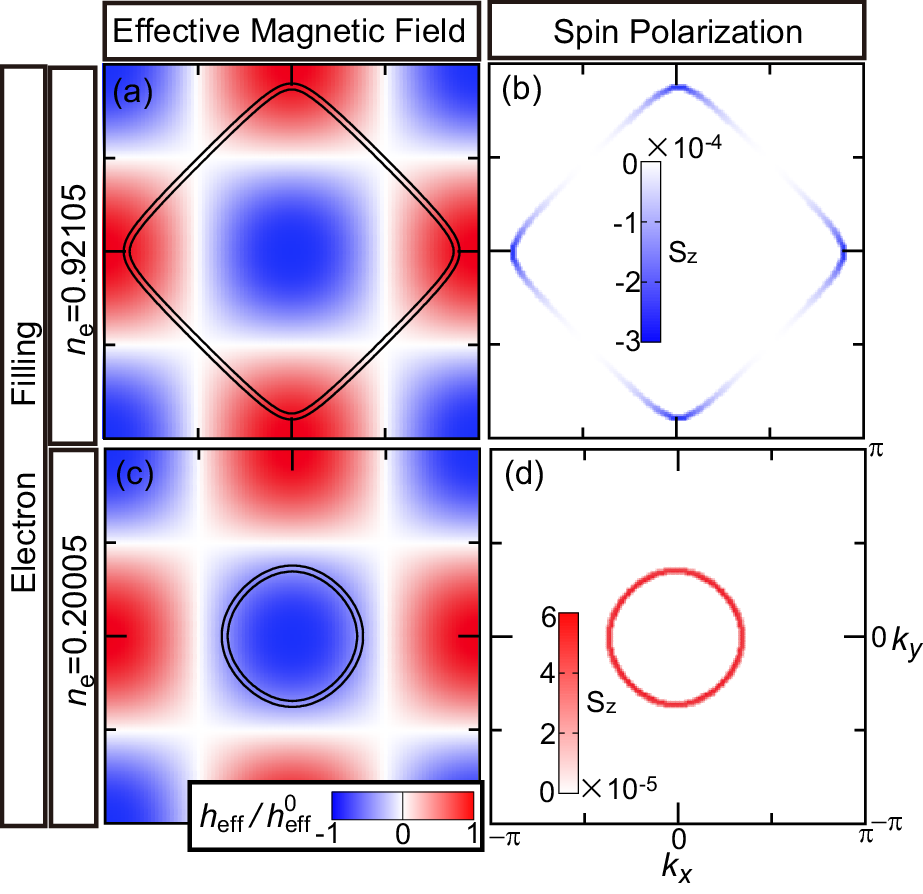}
\caption{(a),(b) Simulated momentum profiles of (a) the effective magnetic field $h^{\rm eff}_{\bm k}$ and (b) contributions to the spin polarization induced by circularly polarized light when the electron filling is relatively high as $n_{\rm e}$=0.92105. (c),(d) Those when the electron filling is relatively low as $n_{\rm e}$=0.20005. The parameters values are $\alpha_{\rm R}$=0.1, $\omega$=1, and $E_0$=0.1. This figure is taken and modified from Ref.~\cite{Tanaka2020} {\copyright} 2020 IOP Publishing.}
\label{Fig08}
\end{figure}
Note that the conditions $f(\widetilde{E}_{-})=1$ and $f(\widetilde{E}_{+})=0$ are satisfied at momentum points located inside of the outer Fermi surface and outside of the inner Fermi surface. Therefore, finite contributions to the spin polarization $s^z_{\bm k}(\ne 0)$ come from only the narrow momentum area between the split two Fermi surfaces. Moreover, the sign of $s^z_{\bm k}$ and the sign of $h^{\rm eff}_{\bm k}$ depend on the momentum area on which the Fermi surfaces are located. Because the size of Fermi surface depends on the electron filling, the sign and magnitude of the spin polarization turns out to attain the sensitive filling-dependence. Figure~\ref{Fig08}(a) shows the color plot of the effective magnetic field $h^{\rm eff}_{\bm k}$ in the momentum space, which is proportional to $-\cos (k_x)\cos (k_y)$, and the Fermi surfaces for the higher electron filling of $n_{\rm e}=$0.92105. In this case, the sign of $h^{\rm eff}_{\bm k}$ at momentum points on the Fermi surfaces are positive as indicated by red color. Consequently, the negative spin polarization $s^z_{\bm k}(<0)$ is induced in the area between the two Fermi surfaces as shown in Fig.~\ref{Fig08}(b). In contrast, when the electron filling is lower as $n_{\rm e}=$0.20005, the Fermi surfaces are in the momentum area for which the effective magnetic field $h^{\rm eff}_{\bm k}$ is negative as indicated by blue color in Fig.~\ref{Fig08}(c). Consequently, the positive spin polarization $s^z_{\bm k}(>0)$ is induced in the area between the two Fermi surfaces as shown in Fig.~\ref{Fig08}(d). This result thoroughly explains the filling-dependent sign and magnitude of the spin polarization observed in the numerical simulations.

Concerning the relevance of the present theory to real materials, n-type semiconductors with large Rashba spin-orbit coupling due to their crystal structures without spatial inversion symmetry are candidate materials with small eletron filling. BiTeI whose conduction band exhibits a large Rashba splitting of $\sim 0.4$ eV is a typical example~\cite{Ishizaka2011,JSLee2011}. In Ref.~\cite{Tanaka2020}, the relevance is discussed in detail. On the contrary, noncentrosymmetric metallic compounds with large Rashba spin-orbit coupling are candidate materials for relatively large electron filling ($n_e$$\sim$0.1-1). Li$_2$Pd$_3$B and Li$_2$Pt$_3$B~\cite{HQYuan2006,KWLee2005}, as well as La$T$Ge$_3$ ($T$=Fe, Co, Rh, Ir) and PrCoGe$_3$~\cite{Kawai2008} are typical examples.

To summarize this chapter, we have discussed recent theoretical studies on the inverse Faraday effect, i.e., photoinduction of spin polarization with circularly polarized light, in Rashba electron systems. Numerical simulations using the tight-binding model with Rashba spin-orbit interaction show that the light electric field induces spin polarization with overwhelmingly high efficiency as compared to the light magnetic field. In fact, the light electric field does not contribute at all to the induction of spin polarization when the spin-orbit interaction is absent. On the contrary, when the pin-orbit interaction is present, the light electric field produces a contribution that is two to three orders of magnitude larger than that of the light magnetic field. The mechanism behind this phenomenon is that the overwhelmingly strong interaction between the light electric field and electron charges can contribute to the induction of the spin polarization through the strong coupling between electron momenta and electron spins mediated by the spin-orbit interactions. In addition to the dependence of the spin polarization on intensity and frequency of light (i.e., $M$$\propto$$E_0^2/\omega^3$), which has been known even in the previous phenomenological theories, the numerical simulations based on the microscopic model have revealed previously unknown important properties, that is, the sign and magnitude of the induced spin polarization depend sensitively on the electron filling. We have also discussed the first modern theory of the inverse Faraday effect based on the Floquet theorem. The theory has turned out to explain that the interplay between the spin-orbit interaction and the circularly polarized light electric field results in the induction of effective static magnetic field perpendicular to the polarization plane. Moreover, the above-mentioned dependencies and properties found in the numerical simulations have been fully reproduced. Not only the Rashba electron systems but also any systems with strong spin-orbit interactions provide a unique platform for realizing the photoinduction of spin polarization of high efficiency. We hope that research on the photoinduced spin polarization taking advantage of the spin-orbit interactions will develop in the future.

\section{Electromagnon excitations in multiferroics and their intense excitation effects}
Multiferroics originally referred to materials in which multiple ferroic orders such as ferromagnetism, ferroelectricity and ferroelasticity coexist~\cite{Schmid1994}. Recently, however, the term is often used to refer to materials in which magnetism and ferroelectricity coexist, i.e., concurrently magnetic and ferroelectric materials in a broader sense. In such materials, nontrivial cross-correlation responses such as magnetization induced by electric field or electric polarization induced by magnetic field through coupling between spins and electric polarizations called magnetoelectric coupling appear. In fact, many multiferroic materials have been known for a long time, but in those materials, magnetism and ferroelectricity coexist accidentally, and, hence, the magnetoelectric coupling~\cite{PCurie1894} is very weak. No significant cross-correlation response was observed in these multiferroic materials. For this reason, research for the multiferroic materials have not been developed so much.

This situation has changed in 2003 with the discovery of multiferroic nature in the rare-earth perovskite manganites $R$MnO$_3$~\cite{Kimura2003}. In ordinary ferroelectrics, the inversion symmetry of the crystal structure is broken originally. On the contrary, in the newly discovered multiferroic rare-earth manganites with $R$=Tb, Dy and Eu$_{1-x}$Y$_x$~\cite{Kimura2003,Goto2004,Kimura2005,Yamasaki2007}, the spiral order of Mn spins breaks the inversion symmetry and induces the ferroelectric polarization~\cite{Katsura2005,Kenzelmann2005,Arima2006,Yamasaki2008}. They therefore exhibit dramatic cross-correlation properties due to strong magnetoelectric coupling. Subsequently, similar spiral-magnetism-induced multiferroelectric materials have been discovered successively, and their astonishing cross-correlation phenomena have been subjects of a major research field~\cite{Tokura2014,Tokura2006,Tokura2007,SWCheong2007,Khomskii2009,Kimura2007,Seki2010}.

\begin{figure}[tb]
\centering
\includegraphics[scale=0.5]{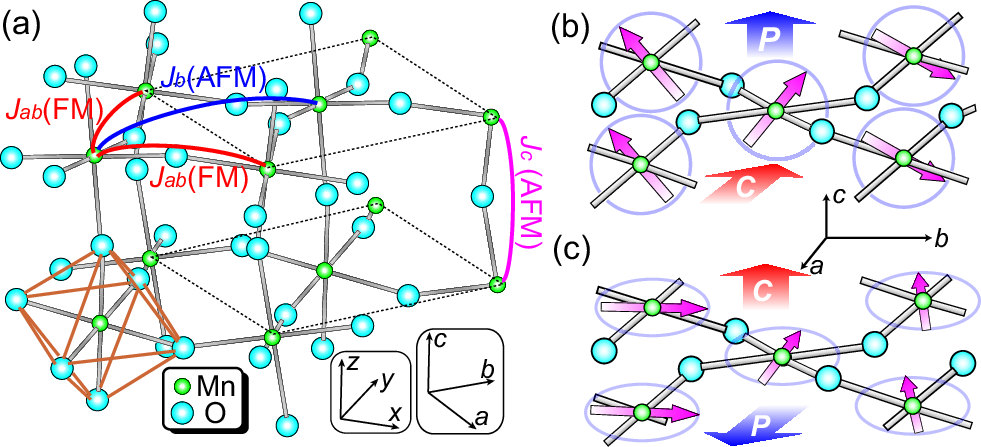}
\caption{(a) Crystal structure and spin exchange interactions in $R$MnO$_3$ with the nearest-neighbor ferromagnetic interaction $J_{ab}$ and next-nearest-neighbor antiferromagnetic interaction $J_b$ frustrated in the $ab$ plane. The inter-plane interaction $J_c$ is antiferromagnetic. (b) In the $bc$-plane spiral magnetism with a vector spin chirality oriented in the $a$-axis, the ferroelectric polarization oriented in the $c$ axis is induced via the antisymmetric exchange striction mechanism.}
\label{Fig09}
\end{figure}
This spiral spin order is stabilized by frustration between the nearest-neighbor ferromagnetic exchange interaction and the next-nearest-neighbor antiferromagnetic exchange interaction in the $ab$ plane shown in Fig.~\ref{Fig09}(a). The physical mechanism of ferroelectricity is explained by the antisymmetric exchange striction mechanism expressed by the following equation~\cite{Katsura2005,Sergienko2006,Mostovoy2006},
\begin{equation}
\label{eq:Eq67}
\bm p_{ij} \propto \bm e_{ij} \times (\bm S_i \times \bm S_j).
\end{equation}
This equation indicates that in the presence of two mutually canted spins $\bm S_i$ and $\bm S_j$, an electric polarization $\bm p_{ij}$ proportional to the vector spin chirality $\bm C_{ij}$=$\bm S_i$$\times$$\bm S_j$ is induced through the spin-orbit interaction. Here $\bm e_{ij}$ is a normalized directional vector connecting the two spin sites. As a result, spiral magnetism as a series of canted spins can induce ferroelectric polarization. In this case, the direction of the ferroelectric polarization depends on the orientation of the spiral spin plane. As can be seen in Fig.~\ref{Fig09}(b) [Fig.~\ref{Fig09}(c)], the $bc$-plane [$ab$-plane] spiral spin order induces ferroelectric polarization along the $c$ axis [$a$ axis]. The orientation of the spin-spiral plane in $R$MnO$_3$ is determined by a subtle competition between magnetic anisotropy and the Dzyaloshinskii-Moriya interaction. For example, the ground-state spin order in Eu$_{1-x}$Y$_x$MnO$_3$ ($x$$\sim$0.5) is the $ab$-plane spiral magnetism, whereas those in TbMnO$_3$ and DyMnO$_3$ are the $bc$-plane spiral magnetism.

In the multiferroic materials, the possibility of spin-wave excitations called electromagnons activated with light electric field via the magnetoelectric coupling has been proposed both theoretically~\cite{Smolenski1982,Katsura2007} and experimentally~\cite{Pimenov2006a,Pimenov2006b}. Electromagnons are novel complex excitations of coupled magnetic and electric dipoles, which are of importance from the viewpoint of fundamental science. These elementary excitations of interest from the viewpoint of technical applications such as tunable electricity and optical filters. Furthermore, the strong coupling between the light electric field and the electric polarizations enables us to achieve intense excitations of electromagnons, which cannot be achieved for conventional magnons excited by light magnetic field. For the intensely excited electromagnons, nonlinear dynamics and cooperative phenomena are expected. It can open up a new field of magnon physics with dramatic phenomena such as magnon Bose-Einstein condensations, light-induced magnetic phase transitions, magnon lasers and magnon Cherenkov radiation as its research targets.

\begin{figure}[tb]
\centering
\includegraphics[scale=0.5]{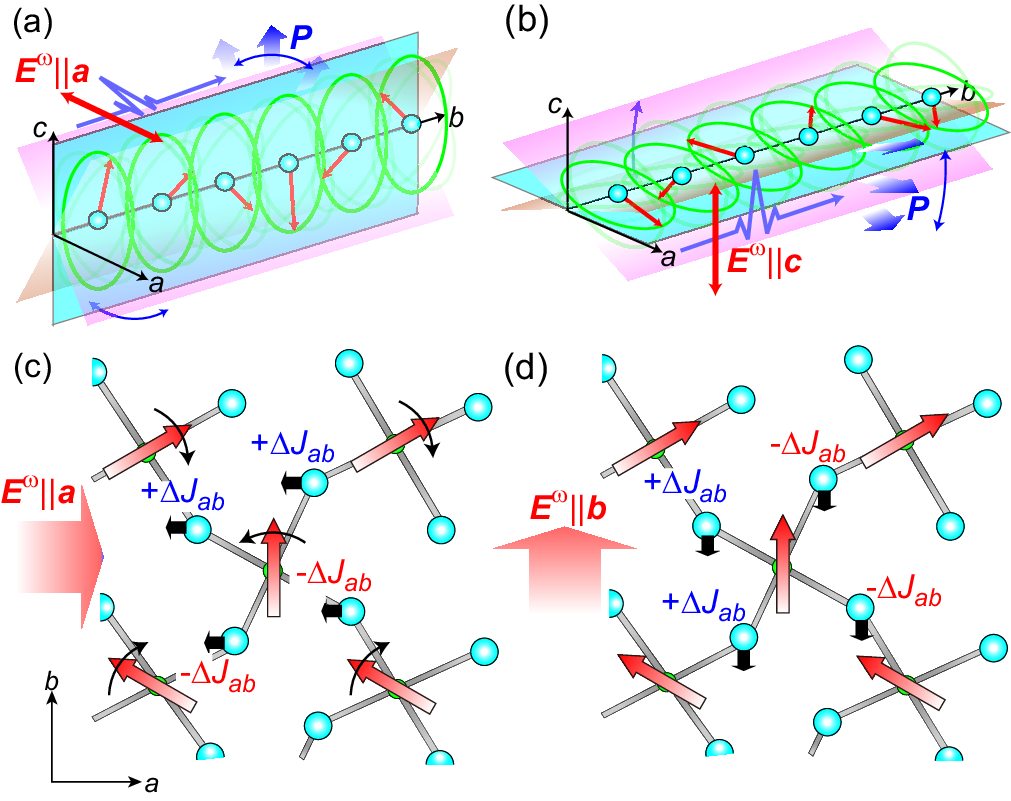}
\caption{(a),~(b) Originally proposed physical mechanism of the electromagnon excitation in $R$MnO$_3$. The ac electric field of light induces an oscillation of the ferroelectric polarization and thus an oscillation of the spin-spiral plane. This mechanism can excite such a magnon mode only when the light electric field is perpendicular to the spin-spiral plane. Namely, the light polarization $\bm E^\omega$$\parallel$$\bm a$ can excite this rotation mode for the $bc$-plane spiral spin order as in (a), whereas the light polarization $\bm E^\omega$$\parallel$$\bm c$ can excite this rotation mode for the $ab$-plane spiral spin order as in (b). These selection rules contradict experimental observations. (c),~(d) Subsequently proposed excitation mechanism. The light electric field couples with local electric polarizations on the zigzag Mn-O-Mn bonds with oxygen anions displaced from the center-of-mass positions of the Mn cations. Through this coupling, the light electric field dynamically modulates the nearest-neighbor ferromagnetic exchange interactions in the $ab$ planes in an oscillatory manner. This mechanism can excite the electromagnon mode only when the light electric field is parallel to the $a$ axis as in (c), which is consistent with the experimental observation. Modulations of the nearest-neighbor ferromagnetic exchange interactions when the light electric field is parallel to the $b$ axis cannot excite cooperative rotational spin oscillations as in (d).}
\label{Fig10}
\end{figure}
A physical mechanism of the electromagnon excitations proposed in the early stage was that the light electric field shakes the ferroelectric polarization induced by the spiral magnetism~\cite{Katsura2007}, which causes a rotational oscillation of the spin-spiral plane (rotational mode) around the spiral propagation axis ($b$ axis) shown in Figs.~\ref{Fig10}(a) and (b). For this mechanism, there necessarily appear selection rules of light polarizations that can excite the electromagnon via this mechanism or those of relative relationship between the light electric-field polarization and the spiral-plane orientation (or the ferroelectric-polarization direction). Specifically, when the spin-spiral plane is the $bc$ plane as shown in Fig.~\ref{Fig10}(a), only the light electric field parallel to the $a$ axis can induce the rotational oscillation of the spin-spiral plane through shaking the ferroelectric polarization $\bm P$$\parallel$$\bm c$. In contrast, when the spin-spiral plane is the $ab$ plane with $\bm P$$\parallel$$\bm a$ as shown in Fig.~\ref{Fig10}(b), the light electric field is required to be parallel to the $c$ axis to induce the rotational oscillation of the spin-spiral plane. In other words, such a magnon mode is expected to be excited only when the light electric field is perpendicular to the spin-spiral plane and the ferroelectric polarization. However, in real experiments for $R$MnO$_3$, resonant optical absorptions associated with electromagnon excitations were observed only when the light electric field is parallel to the $a$ axis, irrespective of whether the spin-spiral plane is $ab$ or $bc$ plane, which is apparently inconsistent with the naive expectation from the proposed mechanism~\cite{Kida2009,Kida2008a,Kida2008b,Takahashi2008,Takahashi2009}.

A new physical mechanism of the electromagnon excitations was then proposed~\cite{Mochizuki2010a,MiyaharaCD,Aguilar2009}. For this mechanism, the light electric field couples to local electric polarizations which are inherently present on the zigzag Mn-O-Mn bonds in the orthorhombically distorted perovskite-type crystal structure called GdFeO$_3$-type structure. The Mn-O-Mn bond angles are reduced from 180$^\circ$ and have no local spatial inversion symmetry. These local polarizations originate from the oxygen anions, each of which are displaced from the center-of-mass position of two Mn cations on each Mn-O-Mn bond. They cancel out in total and do not contribute to the ferroelectric polarization at equilibrium, but their cooperative oscillations can give rise to the electromagnon excitations. 

Specifically, as shown in Figs.~\ref{Fig10}(c) and (d), the divalent oxygen anions O$^{2-}$ are displaced from the original position under application of electric field along the direction indicated by the thick arrow. These anion displacements increase or decrease the Mn-O-Mn bond angles on the respective bonds. Importantly, the strength of the nearest-neighbor ferromagnetic exchange interactions $J_{ab}$ in $R$MnO$_3$ depend sensitively on the Mn-O-Mn bond angle. The ferromagnetic interactions become stronger ($-\Delta J_{ab}$) as the bond angle increases to approach 180$^\circ$, while it becomes weaker ($+\Delta J_{ab}$) as the bond angle decreases. Figures~\ref{Fig10}(c) and (d) show the modulations of the exchange interactions when the light electric field is oriented in the $+a$ and $+b$ directions, respectively. Note that a reversal of the electric field in sign also reverses the sign of $\pm \Delta J_{ab}$. These figures show that the way of the modulations depends on the light polarization or the electric-field direction. 

The modulations of the exchange interactions cause slight rotation of the Mn spins. Specifically, two Mn spins connected by a Mn-O-Mn bond tend to decrease (increase) their angle to approach a parallel (antiparallel) configuration when the electric field increases (decreases) the bond angle and enhances (suppresses) the ferromagnetic interaction. In this way, oscillating light electric field can induce rotational oscillations of the Mn spins. Importantly, only the modulations of the exchange interactions caused by the light electric field applied along the $a$ axis can induce the cooperative rotational oscillation of the spins as shown by thin black arrows in Fig.~\ref{Fig10}(c). This is consistent with the experimental fact that electromagnon excitations are observed only when the light polarization with $\bm E^\omega$$\parallel$$\bm a$.

To confirm this idea, the optical absorption spectra and magnon dispersion relations were calculated through simulating spatiotemporal dynamics of the Mn spins~\cite{Mochizuki2010a,Mochizuki2010b,Mochizuki2011a}. For the simulations, a precise lattice spin model describing the Mn-spin system in $R$MnO$_3$ was used~\cite{Mochizuki2024b,Mochizuki2009a,Mochizuki2009b,Mochizuki2010c,Mochizuki2011b}. To simulate the spin dynamics, the Landau-Lifshitz-Gilbert equation was solved numerically using the Runge-Kutta method. The equation is given by,
\begin{equation}
\label{eq:Eq68}
\frac{\partial \bm S_i}{\partial t}=-\bm S_i \times \bm H^{\rm eff}_i
+ \frac{\alpha_{\rm G}}{S} \left(\bm S_i \times \frac{\partial \bm S_i}{\partial t}\right).
\end{equation} 
The effective magnetic field $\bm H^{\rm eff}_i$ in this equation is obtained by differentiating the Hamiltonian by $\bm S_i$ as $\bm H^{\rm eff}_i = - \partial \mathcal{H} / \partial \bm S_i$. This Hamiltonian is a classical Heisenberg model that treats Mn spins as classical vectors and considers on-site magnetic anisotropies, Dzyaloshinskii-Moriya interactions, frustrated spin exchange interactions, and interactions originating from spin-phonon coupling called biquadratic interactions. These interactions and anisotropies are very small compared to spin-exchange interactions and thus they rarely play important roles in physical phenomena in magnets usually. However, in frustrated magnets like $R$MnO$_3$, the energy scale of the exchange interactions is effectively small, which increases the relative importance of other tiny interactions and anisotropies, which enables them to play decisive roles in physical phenomena. These model parameters were evaluated using perturbation theory and other microscopic calculations. The obtained parameters turned out to accurately reproduce the spiral magnetic orders realized in Eu$_{1-x}$Y$_x$MnO$_3$, TbMnO$_3$ and DyMnO$_3$ in terms of the orientation of the spin-spiral plane and period of the spin spiral.

We simulate the spin dynamics after applying a short-time pulse of light electric field. The effect of this light-pulse irradiation is taken into account through modulating the nearest-neighbor ferromagnetic interactions as shown in Figs.~\ref{Fig10}(c) and (d). We next describe how the optical absorption spectrum can be calculated from the obtained spin-dynamics data. The spin oscillations excited by the light pulse induce oscillations of electric polarizations through the magnetostriction mechanism. As argued above, when the oxygen ions on the Mn-O-Mn bonds are displaced, the nearest-neighbor ferromagnetic interactions are modulated accordingly, which change the relative angles between the Mn spins. Inversely, when the relative spin angles is decreased (increased), the oxygen ions spontaneously shift to enhance (reduce) the nearest-neighbor ferromagnetic interactions via increasing (decreasing) the Mn-O-Mn bond angle in order to increase (decrease) the energy gain (cost) due to the exchange interactions. Consequently, a decrease or increase of the local electric polarization scaled with the inner product of spins, $\bm S_i \cdot \bm S_j$, occurs on each Mn-O-Mn bond, and a dynamical component of the net electric polarization $\bm P_{\rm s}(t)$ appears, which is described by,
\begin{equation}
\label{eq:Eq69}
\bm P_{\rm s}(t)=\sum_{i,j} \bm \Pi_{ij} \; \bm S_i(t) \cdot \bm S_j(t).
\end{equation}
This mechanism is called the symmetric exchange strain mechanism. Here $\bm \Pi_{ij}$ is a form factor which reflects the double period of the zigzag Mn-O-Mn bond. In this way, the time evolution of the dynamical electric polarization $\bm P_{\rm s}(t)$, which appears as a response to the light electric-field pulse, can be calculated from the simulated time profiles of Mn spins. By performing the Fourier transformation, the $\omega$-dependence of the dielectric coefficient $\varepsilon$ can be calculated. Its imaginary part Im$\varepsilon(\omega)$ corresponds to the experimentally observed optical absorption spectrum.

\begin{figure}[tb]
\centering
\includegraphics[scale=0.5]{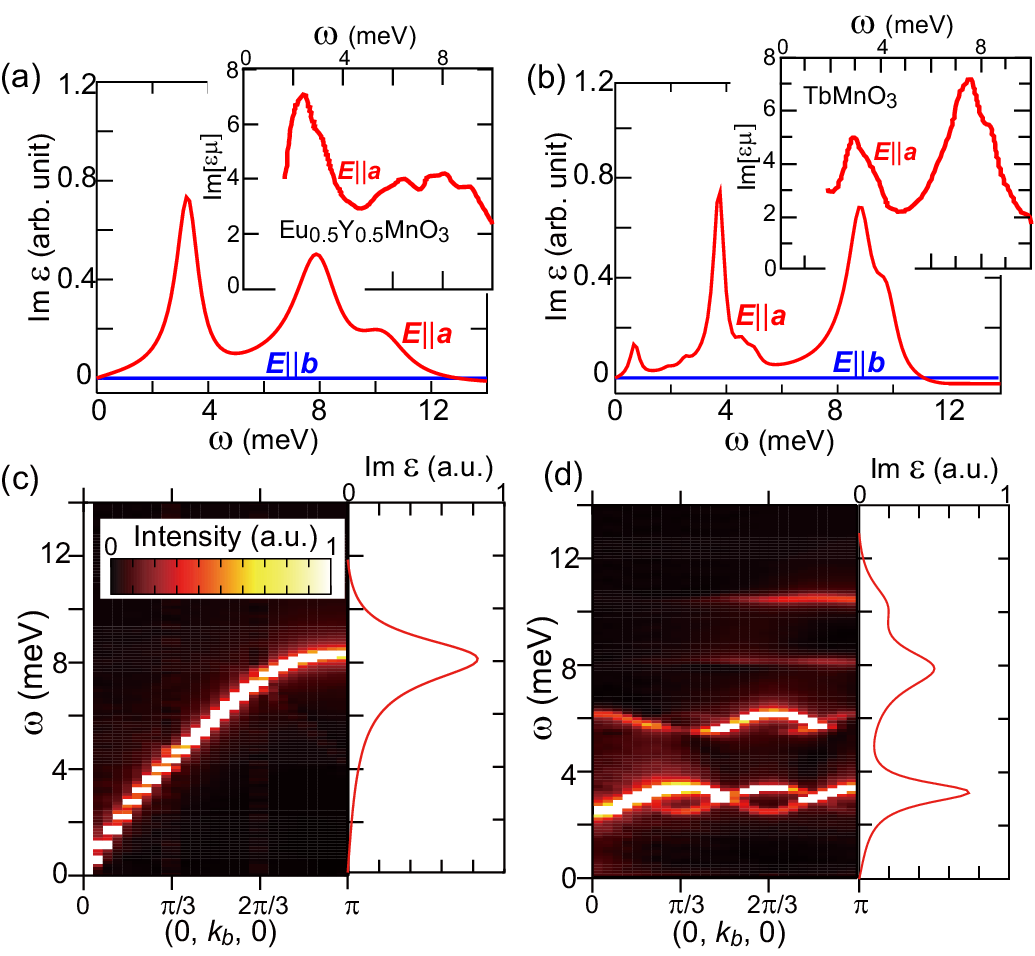}
\caption{(a),(b) Calculated optical absorption spectra of (a) Eu$_{1-x}$Y$_x$MnO$_3$ with $ab$-plane spin cycloid and (b) TbMnO$_3$ with $bc$-plane spin cycloid. Both spectra have two remarkable peaks at terahertz frequencies only when the light electric field is parallel to the $a$ axis. Insets show the experimental spectra. (c),(d) Calculated magnon dispersion relations (left) and optical absorption spectra (right) for (c) Eu$_{1-x}$Y$_x$MnO$_3$ with $ab$-plane spin cycloid and (d) TbMnO$_3$ with $bc$-plane spin cycloid. This figure is taken and modified from Ref.~\cite{Mochizuki2010a} {\copyright} 2010 American Physical Society.}
\label{Fig11}
\end{figure}
Figures~\ref{Fig11}(a) and (b) show the calculated optical absorption spectra of Eu$_{1-x}$Y$_x$MnO$_3$ ($ab$-plane spiral magnetism) and TbMnO$_3$ ($bc$-plane spiral magnetism). Noticeably, the spectra exhibit two characteristic peaks in the terahertz regime irrespective of the orientation of spin-spiral plane, only when the light electric field is parallel to the $a$ axis, i.e., $\bm E^\omega$$\parallel$$\bm a$. These theoretical spectra as well as their selection rule in terms of the light polarization reproduce well the experimental results shown in the insets. Of the two peaks in the spectra, the higher-frequency peak at $\sim$2 THz ($\sim$8 meV) is attributed to the magnon mode indicated by the thin black arrows in Fig.~\ref{Fig10}(c). This is the zone-edge magnon mode corresponding to the wavenumber ($k_a$, $k_b$, $k_c$)=(0, $\pi$, 0). In contrast, another peak at $\sim$0.7 THz ($\sim$3-4 meV) is attributed to higher- harmonic magnon modes. In fact, the rotation of Mn spins in the spiral magnetism of $R$MnO$_3$ is not uniform and contains large higher harmonic components due to the site-dependent magnetic anisotropies and the staggered Dzyaloshinskii-Moriya interactions. Therefore, magnon folding occurs at wavenumbers corresponding to the higher harmonic components, which gives rise to peaks on the low energy region. In fact, the uniformly rotating spiral magnetism obtained from a Hamiltonian containing only the spin-exchange interactions yields a single magnon dispersion and an optical absorption spectrum with a single peak originating from the zone-edge magnon as shown in Fig.~\ref{Fig11}(c). In contrast, for the spiral magnetism with higher harmonic components obtained from a Hamiltonian that incorporates all the interactions of the real materials, folding and cross-repulsions can be seen in the magnon dispersion relations, and, consequently, the low-frequency peak appears in addition to the high-frequency peak as shown in Fig.~\ref{Fig11}(d).

Now the mechanism of the electromagnon excitations in multiferroic $R$MnO$_3$ has been fully elucidated. Then, our next interest is possible interesting phenomena expected for intensely excited electromagnons with strong laser irradiation. As an example, we introduce a theoretically proposed light-induced spin-chirality switching by means of intense electromagnon excitations~\cite{Mochizuki2010b,Mochizuki2011a}. Chirality is a universal and important concept that appears not only in physics but also in broad areas of science such as biology and chemistry. Functions of biological proteins are governed by chirality of the structure, and the properties of many drugs and chemicals are inseparably linked to chirality of their crystals. How to control chirality has been a longstanding challenge of mankind. Indeed, the method of asymmetric synthesis in chemistry is regarded as an epoch-making discovery. Here we argue that the chirality of spins can be controlled and/or manipulated at picosecond speeds by light irradiation.

\begin{figure}[tb]
\centering
\includegraphics[scale=0.5]{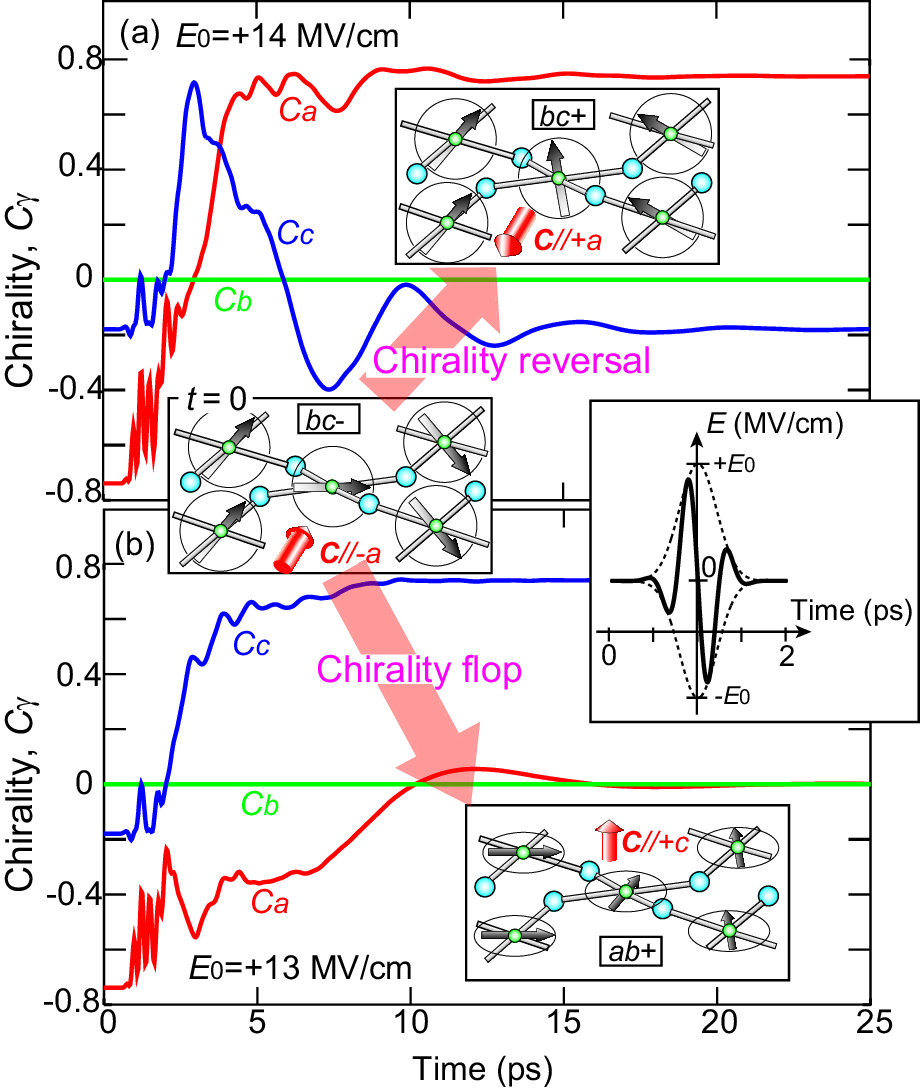}
\caption{Simulated time profiles of the $a$-, $b$- and $c$-axis components of the spin chirality after application of an intense light pulse. (a) 180$^\circ$ reversal of the vector spin chirality from $\bm C$$\parallel$$-\bm a$ to $\bm C$$\parallel$$+\bm a$ for $E_0$=+14 MV/cm. (b) 90$^\circ$ flop of the vector spin chirality from $\bm C$$\parallel$$-\bm a$ to $\bm C$$\parallel$$+\bm b$ for $E_0$=+13 MV/cm. Inset shows the temporal waveform of the applied light pulse. This figure is taken and modified from Ref.~\cite{Mochizuki2010b} {\copyright} 2010 American Physical Society.}
\label{Fig12}
\end{figure}
A sinusoidally oscillating light electric-field pulse with a Gaussian envelope parallel to the $a$ axis (i.e., $\bm E$$\parallel$$\bm a$) is applied to the ground-state $bc$-plane spiral spin order in TbMnO$_3$. The temporal waveform of the pulse is given by,
\begin{align}
\label{eq:Eq70}
E(t)=-E_0 \sin(\omega t) \; \exp\left[-\frac{(t-t_0)^2}{2\sigma^2}\right].
\end{align}
Here the frequency $\omega$ is fixed at 2.1 THz, which corresponds to the resonance frequency of the higher-lying electromagnon excitation in TbMnO$_3$, while the Gaussian half width $\sigma$ is fixed at 0.5 ps. The spin dynamics after the pulse application is simulated using the LLG equation. Figures~\ref{Fig12}(a) and (b) show the simulated time profiles of the $a$-, $b$- and $c$-axis components of the vector spin chirality $\bm C$. When the field strength is $E_0$=+14 MV/cm, the sign of $C_a$, which originally has a negative value, is overturned to have a positive value as seen in Fig.~\ref{Fig12}(a). This indicates that a 180$^\circ$ chirality reversal from the clockwise $bc$-plane spiral with $C_a(<0)$ to the counterclockwise $bc$-plane spiral with $C_a(>0)$ takes place. In contrast, as seen in Fig.~\ref{Fig12}(b), when the field strength is $E_0$=+13 MV/cm, the $a$-axis component $C_a$, which originally has a large negative value, becomes zero, while the $c$-axis component $C_c$, which is originally almost zero, turns to have a large positive value. This indicates that a 90$^\circ$ chirality flop from the clockwise $bc$-plane spiral with $C_a(<0)$ to the counterclockwise $ab$-plane spiral with $C_c(>0)$ takes place. These reversal and flop of the chirality occur very quickly typically within a few picoseconds.

\begin{figure}[tb]
\centering
\includegraphics[scale=0.5]{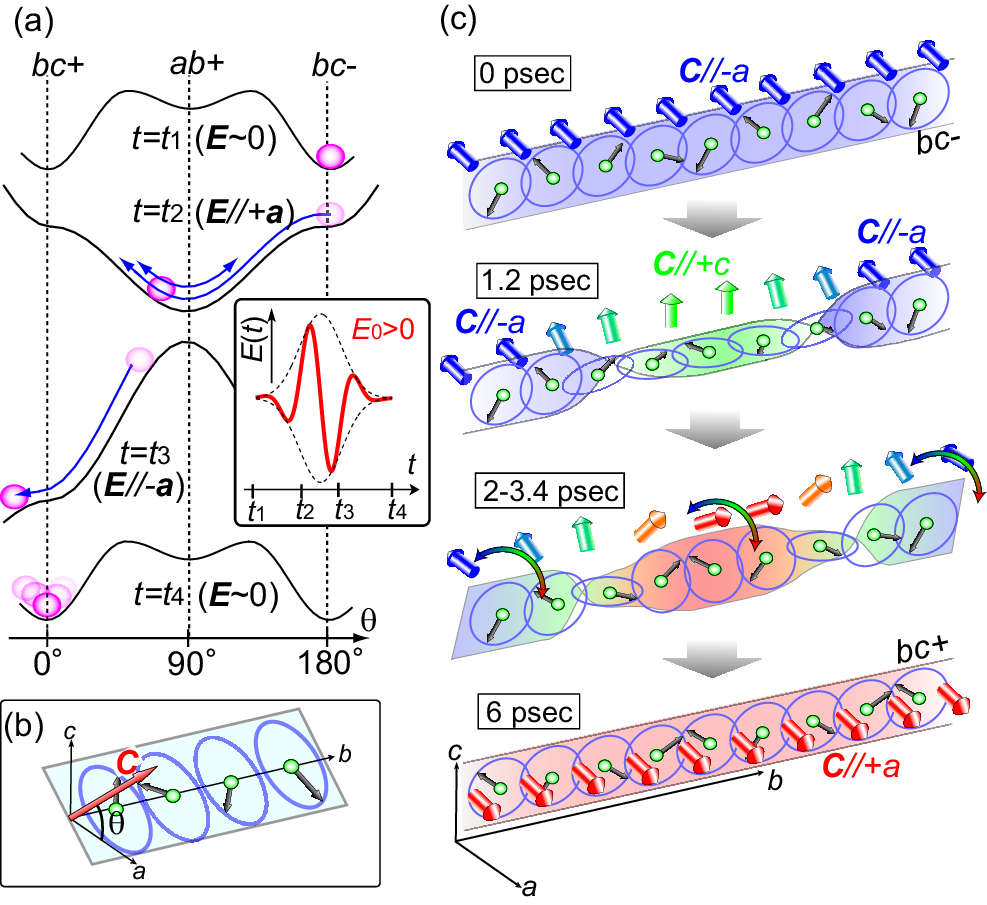}
\caption{(a) Schematics of temporal modulation of the potential structure and dynamical process of the chirality reversal induced by light-pulse irradiation. (b) Definition of the angle $\theta$ of the vector spin chirality. (c) Dynamical spatial patterns emerging during the transient process of chirality reversal. This figure is taken and modified from Ref.~\cite{Mochizuki2010b} {\copyright} 2010 American Physical Society.}
\label{Fig13}
\end{figure}
These spin-chirality reversal and flop are caused by dynamic modulation of potential structure induced by the oscillating electric field of light and inertial motion of the spin-spiral plane that acquires effective mass due to magnetic anisotropy. Under application of the light electric field $\bm E$$\parallel$$\bm a$, the nearest-neighbor ferromagnetic exchange interactions in the $ab$ plane increase or decrease as shown in Fig.~\ref{Fig10}(c), which results in modulations of the relative angles between neighboring Mn spins. Specifically, the relative spin-spin angles become smaller (larger) on the Mn-O-Mn bonds with stronger (weaker) ferromagnetic exchange interactions. These spin-spin angle modulations result in modulations of the local spin-chirality vectors $\bm C_{ij}$=$\bm S_i$$\times$$\bm S_j$ defined between the nearest neighbor Mn spins $\bm S_i$ and $\bm S_j$. The way of the modulations depends on the rotation sense of spin spiral (clockwise and counterclockwise) or the sign of the net vector spin chirality $\bm C$$\propto$$\sum_{<i,j>}$$\bm C_{ij}$. Because of the presence of Dzyaloshinskii-Moriya interactions associated with the cross-products of spins $\bm S_i$$\times$$\bm S_j$, the spiral spin order has an energy that depends on the direction of spin chirality or the orientation of spin-spiral plane under application of electric field. When an electric field $\bm E$$\parallel$$+\bm a$ ($\bm E$$\parallel$$-\bm a$) is applied, the energy of the counterclockwise $ab$-plane spin spiral decreases (increases), while the energy of the $bc$-plane spin spiral remains unchanged irrespective of its rotation sense. Consequently, the potential structure is dynamically modulated by the oscillating electric field of light [inset of Fig.~\ref{Fig13}(a)]. Figure~\ref{Fig13}(b) shows the dynamical modulation of the potential structure as a function of $\theta$ where $\theta$ is an angle between the net chirality vector $\bm C$ and the $a$ axis [Fig.~\ref{Fig13}(b)].

Unlike the usual polarization reversals under a static electric field, the light-induced chirality reversal is expected to exhibit interesting phase-transition dynamics involving the formation of dynamical spatial patterns in the transient process. Specifically, dynamical stripes of chirality domains are expected to be formed [Fig.~\ref{Fig13}(c)]. When the clockwise $bc$-plane spiral spin state is irradiated by a light pulse, stripes of chirality domains appear. The chirality vector in each domain oscillates in an opposite phase to those in the neighboring domains. Through this oscillatory chirality-domain stripe state, the system is settled into the counterclockwise $bc$-plane spiral spin state with a uniform chirality vector pointing in the direction opposite to the original one.

To realize the chirality reversal and flop, the frequency of light should be tuned at the frequency of the higher-lying electromagnon mode [Fig.~\ref{Fig11}(b)] to efficiently excite this mode. An intense light electric field ($E_0$$\sim$10 MV/cm in the present case) is also required, but the stronger light is not necessarily better. Figure~\ref{Fig13}(a) indicates that the chirality oscillation around the potential bottom under the light electric field $E$$\parallel$$+\bm a$ at time $t$$\sim$$t_2$ and the timing of the sign change of the light electric field at time $t$=$t_3$ should be well synchronized such that the chirality falls into an another potential minimum. If the timing is nicely tuned, the chirality can be stopped in the $ab$-plane spiral state to realize the 90$^{\circ}$ chirality flop. The threshold light electric field becomes smaller when the pulse duration is increased. Dramatic phenomena like chirality switching can be achieved by exploiting the resonance nature of the electromagnon excitations and tuning the synchronization and timing of the oscillations. The switching can also be realized with a weaker light electric field by exploiting the assistance from thermal fluctuations and/or by choosing materials located near the phase boundary between the $ab$- and $bc$-plane spiral spin phases. Numerical simulations demonstrated that switching among four distinct chirality states corresponding to two-by-two combinations of different rotation senses (clockwise and counterclockwise) and different spin-spiral planes ($ab$- and $bc$-planes) by tuning parameters of the applied light pulse with respect to sign, shape, length and intensity.

To summarize this section, recent theoretical studies on the electromagnon excitations in multiferroic rare-earth perovskite manganites $R$MnO$_3$ have been discussed. In fact, electromagnons are elementary excitations which can be ubiquitously observed in multiferroic materials. There are several excitation mechanisms other than the mechanism discussed in this section, which are not yet fully understood. However, electromagnons, which can be intensely excited by the strong coupling between the light electric field and the electric polarizations, have high potential as a means of efficient control of magnetism by light, as demonstrated in this section. We hope that research on the photocontrol of magnetism using electromagnons will develop in the future.

\section{Summary}
In this article, we have reviewed recent theoretical studies on the photocontrol of magnetism with high efficiency in spin-charge coupled systems by taking the following three systems as examples, double-exchange magnets, Rashba electron systems, and multiferroic materials. More specifically, the following three topics have been discussed, [1] theoretical study of photoinduced magnetic phase transitions in double-exchange models, [2] microscopic theory of highly efficient photoinduction of spin polarization in Rashba electron systems, [3] theory of electromagnon excitations in multiferroics and predicted their intense excitation effects. In these spin-charge coupled systems, the coupling between spins and charges are mediated, respectively, by exchange interactions between localized spins and conduction-electron spins, spin-orbit interactions between electron spins and electron momenta, and magnetoelectric coupling between spins and electric polarizations. Note that light electromagnetic waves are composed of ac electric-field and ac magnetic-field components. Their strengths are not independent, but a relation $E_0/B_0=c$ holds where $E_0$ ($B_0$) is the amplitude of light electric (magnetic) field and $c$ is the speed of light. Importantly, the interactions between light electric field and electron charges are two to three orders of magnitude larger than the interactions between light magnetic field and electron spins. However, the electron spins or magnetization in materials can couple only with light magnetic field usually so that dramatic effects and phenomena are hardly realized as far as the magnetic light-matter interaction is exploited. On the contrary, in spin-charge coupled magnets, the strong electric light-matter coupling can be converted to an efficient drive of electron spins or magnetization. Therefore, the spin-charge coupled systems provide wonderful treasure chests of many dramatic effects, astonishing phenomena, and useful functions in photocontrol of magnetism. This field of research has developed rapidly in recent years along with the development of laser technology. We hope that this article contributes to its development.

\section{Acknowledgment}
I am grateful to Yasuhiro Tanaka for useful discussions. This work was supported by Japan Society for the Promotion of Science KAKENHI (Grants No.20H00337, No.24H02231 and No.25H00611), CREST, the Japan Science and Technology Agency (Grant No. JPMJCR20T1), and Waseda University Grant for Special Research Projects (2024C-153 and 2025C-133).

\end{document}